\documentclass[pdftex,twocolumn,apjfonts,10pt,apj,numberedappendix]{emulateapj}
\usepackage[letterpaper,top=1.875in, bottom=0.14in, left=1.07in, right=0.58in]{geometry}
\usepackage{graphicx}
\usepackage{multirow}

\newcommand{\eg}{\protect{\em e.g.\ }}
\newcommand{\ie}{\protect{\em i.e.\ }}
\newcommand{\cf}{\protect{\em c.f.\ }}
\def\d{\mathrm{d}}

\begin{document}
\title{Deep HST Imaging in 47~Tucanae: A Global Dynamical Model}
\author{J. Heyl$^1$}
\author{I. Caiazzo$^1$}
\author{H. Richer$^1$}
\author{J. Anderson$^2$}
\author{J. Kalirai$^2$}
\author{J. Parada$^1$}
\affil{$^1$ Department of Physics and Astronomy, University of British
  Columbia, Vancouver BC V6T 1Z1 Canada}
\affil{$^2$ Space Telescope Science Institute, Baltimore MD 21218}
\begin{abstract}
Multi-epoch observations with ACS and WFC3 on HST provide a unique and comprehensive probe of stellar dynamics within 47 Tucanae. We confront analytic models of the globular cluster with the observed stellar proper motions that probe  along the main sequence from just above 0.8 to 0.1M$_\odot$ as well as white dwarfs younger than one gigayear.  One field lies just beyond the half-light radius where dynamical models  (\eg lowered Maxwellian distributions) make robust predictions for the stellar proper motions.  The observed proper motions in this outer field show evidence for anisotropy in the velocity distribution as well as skewness; the latter is evidence of rotation.  The measured velocity dispersions and surface brightness distributions agree in detail with a rotating, anisotropic model of the stellar distribution function with mild dependence of the proper-motion dispersion on mass.  However, the best fitting models under-predict the rotation and skewness of the stellar velocities.  In the second field, centered on the core of the cluster, the mass segregation in proper motion is much stronger.  Nevertheless the model developed in the outer field can be extended inward by taking this mass segregation into account in a heuristic fashion.  The proper motions of the main-sequence stars yield a mass estimate of the cluster of $1.31 \pm 0.02 \times 10^6 \mathrm{M}_\odot$ at a distance of 4.7 kpc.  By comparing the proper motions of a sample of giant and sub-giant stars with the observed radial velocities we estimate the distance to the cluster kinematically to be $4.29 \pm 0.47$ kpc.
\end{abstract}
\maketitle

\section{Introduction}
\label{sec:introduction}

Globular clusters are nature's finest laboratory for stellar evolution, star formation and gravitational dynamics, and perhaps the best studied globular cluster is 47~Tucanane. Its proximity to Earth \citep[about four to five kpc, \eg][]{2012AJ....143...50W} along with its richness \citep[about one million stars, \eg][]{2015ApJ...803...29W} make it a prototypical globular cluster. 

The segregation of stars in 47 Tucanae by mass has been the focus of many studies \citep[\eg][]{1982AJ.....87..990D,1996ASPC...92..257A,2005ApJ...625..796H,2015ApJ...815...95Z}. More recently \citet{Heyl14diff} studied the dynamical processes occurring today in the center of the cluster and revealed on-going mass segregation for the first time.  \citet{Heyl14massloss} pinpointed the epoch of mass loss of the stars currently becoming white dwarfs, \citet{2016arXiv160505740P} constrained the masses of post-main-sequence stars and \citet{2016arXiv160902115P} constrained the masses of blue-straggler stars. Here, our primary focus will be the dynamics in a field about two half-light radii from the center of the cluster.  Unlike in the core, dynamical relaxation should have had a minor impact in this field. In fact, \citet{Rich1347tuc} and \citet{2014A&A...568L...4K} found differences in dynamical states of stars among its stellar populations which could be possibly linked to their formation mechanism.

We will examine the dynamics in 47~Tucanae in detail using proper motions as we did also for an outlying field in NGC~6397 \citep{Heyl116397dyn}. Starting with \citet{1960MNRAS.120..463F}, the dynamics of the stars within 47~Tucanae have been studied extensively \citep[\eg][]{1983A&AS...54..495M,1984A&A...134..118M,1986A&A...166..177D,1986A&A...166..122M,1988A&A...191..215M,1989A&A...214..106M,1995AJ....110.1699G,2003AJ....126..772A,2006ApJS..166..249M,2010ApJ...711L.122L,2013ApJ...772...67B,2016A&A...586A..77C}.
Studies of outlying fields complement the work in the cores of the clusters \citep[\eg][]{2014ApJ...797..115B}. In the outlying fields, relaxation times are generally longer, so one expects that the dynamics may reflect the ancient conditions of the cluster.  Furthermore, the angular momentum naturally can play a stronger role in the outskirts of a cluster. From our data in NGC~6397 we constrained the relaxation time to be greater than or about 800~Myr \citep{Heyl116397dyn}, much smaller than the age of the cluster.  We also found that the proper motions and star counts are well characterized by a lowered-isothermal sphere, a Michie-King model \citep{1963MNRAS.125..127M,1966AJ.....71...64K} with mass segregation both in position and proper motion and no evidence of anisotropy.  The only component that did not fit this picture of a relaxed stellar system were the young white dwarfs which did not yet have time to relax \citep[see also][]{2013ApJ...778..104R}. 

Most recently \citet{2017ApJ...844..167B} have used proper motions in several fields to study the dynamics of 47 Tuc. In fact, their central field and calibration field correspond to the core WFC3 and outer ACS field in this paper. Whereas the focus of \citet{2017ApJ...844..167B} is the rapid rotation of the cluster, our goal here is similar to our goal in \citet{Heyl116397dyn}; we will develop a global nearly analytic model of the dynamics and stellar density of 47~Tucanae and focus on the details of the observed distributions of proper motions. We will augment the data here with wide-field photometric data from \citet{1995AJ....109..218T}.  Table~\ref{tab:data} gives dynamical and other data for 47~Tucanae, including some key results from this paper. In the sections that follow, we will outline the observations themselves in \S\ref{sec:observations} and the theoretical model to account for the stellar densities and motions in \S\ref{sec:modelling-47-tucanae}. Because the stars in this field have not yet relaxed, this model is necessarily richer than the one that we used for NGC~6397. It must account for both the rotation and the anisotropic velocity distribution of the cluster. \S\ref{sec:results} examines the dynamics of the stars in the cluster both descriptively and in terms of a model that characterizes the dynamical and photometry comprehensively.  In \S\ref{sec:discussion} we outline the wider implications of these results. 
%  and suggest future directions.

\begin{table*}
\begin{center}
\caption{Basic data on 47 Tucanae}
\label{tab:data}
\begin{tabular}{lll}
\hline \hline
\multicolumn{1}{c}{Property} & 
\multicolumn{1}{c}{Value} & 
\multicolumn{1}{c}{Reference} \\
\hline
Cluster center (J2000) & 
$\alpha=00^\mathrm{h}24^\mathrm{m}05^\mathrm{s}\!\!.71, \delta=-72^\circ04'52''\!\!.7$
& \citealt{2010AJ....140.1830G}\\
Galactic coordinates & $l=305^\circ\!\!.8948, b=-44^\circ\!\!.8893$ &
\citealt{2011AJ....142...66G} \\
Apparent Magnitude & $V_\mathrm{tot} =   3.95$ &
\citealt{1996AJ....112.1487H} \\
%Luminosity & $L_V = 2.9 \times 10^4 d_2^2 \mathrm{L}_\odot$ &
Luminosity & $L_V = 5.0 \times 10^5 d_{4.7}^2 \mathrm{L}_\odot$ &
\citealt{1996AJ....112.1487H} \\
Integrated colors & $B-V=0.88, U-V=1.25$ & \citealt{1996AJ....112.1487H}  \\
Metallicity & $[\mathrm{Fe/H}] =-0.75$ & \protect{\citealt{2014MNRAS.437.3274V}} \\
Foreground reddening & $E(B-V) = 0.04 \pm 0.02$ & \protect{\citealt{2007A&A...476..243S}} \\
Foreground absorption & $A_V = 0.124$ & \citealt{1989ApJ...345..245C} \\
                      & $A_\mathrm{F814W} = 0.07$ & \citealt{2005PASP..117.1049S} \\
Field contamination ($V \leq 21$) & $\Sigma_\mathrm{fore} \approx 1.1$
stars arcmin$^{-2}$ & \citealt{1985ApJS...59...63R}\\
\phantom{Field contamination }($V \leq 29$) & $\phantom{\Sigma_\mathrm{fore}} \approx 9.1$
stars arcmin$^{-2}$ & \\
Structural Parameters: \\
%~~Total Mass$^*$ & $5.6 \pm 0.4  \times 10^4 d_2^3 \mathrm{M}_\odot$ & This
~~Total Mass$^*$ & $\left ( 1.31 \pm 0.02 \right ) \times 10^6 d_{4.7}^3 \mathrm{M}_\odot$ & This
paper, \S~\ref{sec:mass-47-tucanae} \\
~~Core Radius & 22~arcseconds & \citealt{1995AJ....109..218T} \\
~~Half-Light Radius & $R_h=2.9$~arcminutes & \citealt{1995AJ....109..218T} \\
Anisotropic Michie-King Model: & & \\
~~Central Escape Velocity & 2.95 mas/yr & This paper,
\S~\ref{sec:modelling-47-tucanae} \\
% ~~Central Escape Velocity & 2.25 $d_2^{-1}$ mas/yr & Gnedin ? \\
%~~Central Escape Velocity & 2.8 $d_{2.53}^{-1}$ mas/yr & \citealt{1999ApJ...522..935G} \\
~~$\sigma-$parameter  & 0.72 mas/yr & This
paper, \S~\ref{sec:modelling-47-tucanae} \\
~~$r_a-$parameter &  6.1 arcminues &  This
paper, \S~\ref{sec:modelling-47-tucanae} \\
Rotating Lupton-Gunn Model: & & \\
~~Central Escape Velocity & 2.90 mas/yr & This paper,
\S~\ref{sec:modelling-47-tucanae} \\
% ~~Central Escape Velocity & 2.25 $d_2^{-1}$ mas/yr & Gnedin ? \\
%~~Central Escape Velocity & 2.8 $d_{2.53}^{-1}$ mas/yr & \citealt{1999ApJ...522..935G} \\
~~$\sigma-$parameter  & 0.70 mas/yr & This
paper, \S~\ref{sec:modelling-47-tucanae} \\
~~$r_a-$parameter &  5.6 arcminutes &  This
paper, \S~\ref{sec:modelling-47-tucanae} \\
~~$r_b-$parameter &  6.5 arcminutes &  This
paper, \S~\ref{sec:modelling-47-tucanae} \\
Heliocentric distance: & & \\
~~Eclipsing binary & $4.43 \pm 0.17$~kpc &
\protect{\citealt{2010AJ....139..329T}} \\
~~Main-sequence fitting & $4.47$~kpc &
\protect{\citealt{2009AJ....138.1455B}} \\
~~White-dwarf fit             & $4.70 \pm 0.04 \pm 0.13$~kpc & \protect{\citealt{2012AJ....143...50W}} \\
~~Kinematic & $4.1 \pm 0.5$~kpc &
\protect{\citealt{2006ApJS..166..249M}} \\
~~          & $4.15 \pm 0.13$~kpc &
\protect{\citealt{2015ApJ...812..149W}} \\
~~          & $4.29 \pm 0.47$~kpc &
This paper, \S~\ref{sec:centr-prop-moti} \\
Timescales at 6.4': & & \\
~~Crossing time from proper motions & $\tau_c \approx r/{\hat \sigma}_\mu \approx
0.7$~Myr & This paper, \S~\ref{sec:prop-moti-disp} \\
~~Relaxation time & $\tau_r \approx 26~d_{4.7}^6$~Gyr & This paper, \S~\ref{sec:mass-47-tucanae} \\
%% ~~Evaporation time & $\tau_e > \left(d\ln N/dt\right)^{-1} \approx 6$~Gyr & This paper, \S~\ref{sec:stellar-escapers} \\
%% ~~Evaporation time & *$\tau_e = \left(d\ln N/dt\right)^{-1} \approx
%% 3$~Gyr & This paper, \S~\ref{sec:stellar-escapers} \\
%% ~~Relaxation time from white-dwarfs & *$\tau_r \gtrsim 0.7$~Gyr & This
%% paper, \S~\ref{sec:prop-moti-distr} \\
%% ~~Dynamical relaxation time (in our field) & *$\tau_r \approx 1$~Gyr & This paper, \S~\ref{sec:prop-moti-disp} \\
%% ~~Dynamical relaxation time (at $R_h$) & *$\tau_{rh} \approx 0.3$~Gyr & This paper, \S~\ref{sec:modelling-47-tucanae} \\

\hline
   \end{tabular}
\end{center}
\end{table*}

\section{Observations}
\label{sec:observations}

\begin{figure}
\includegraphics[width=\columnwidth]{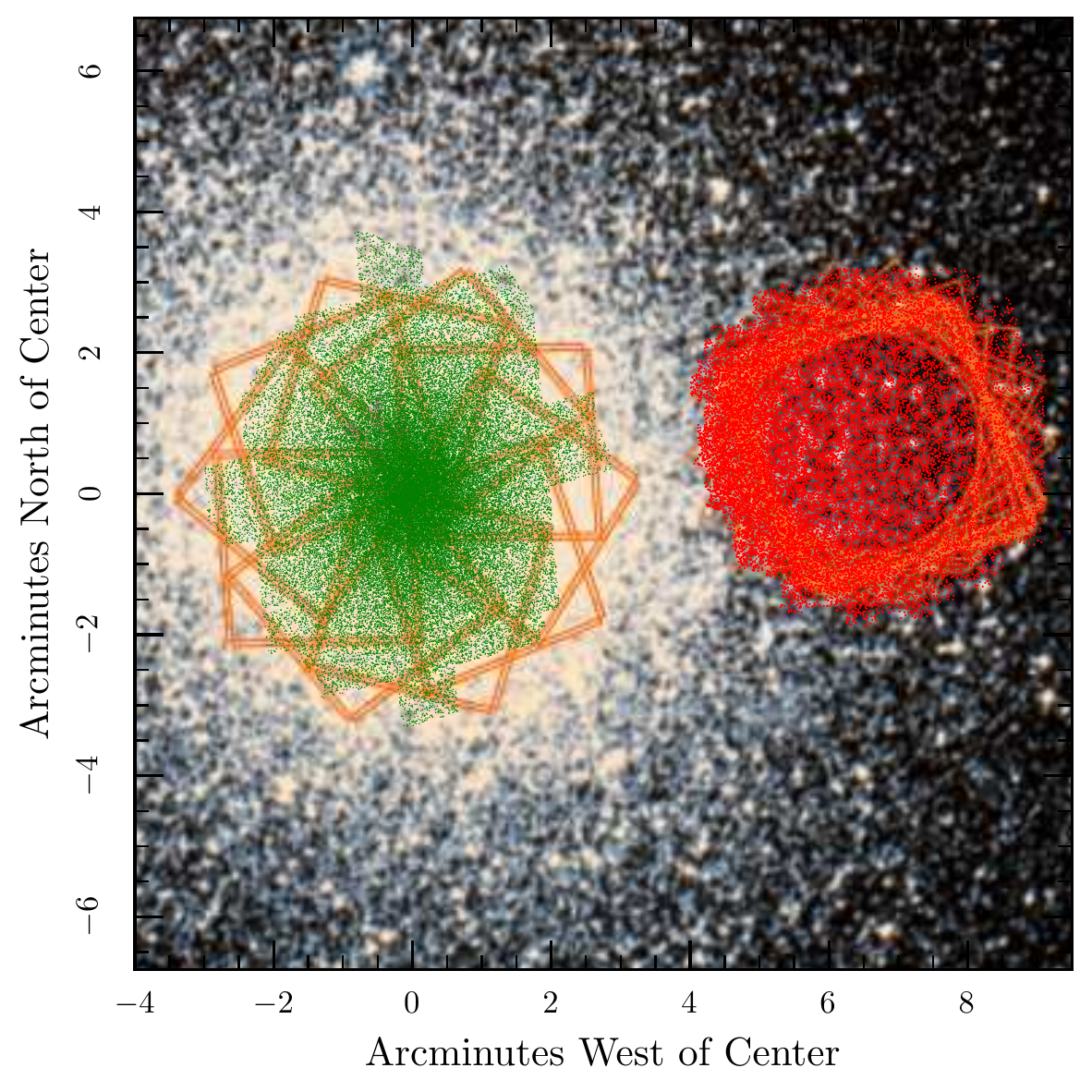}
\caption{The locations of stars with measured proper motions with WFC3 (green on the left) and ACS (red on the right) superimposed on the approximate boundaries of the fields for GO-12971 (left) and GO-11677 (right).}
\label{fig:both_field}
\end{figure}

\begin{figure}
\includegraphics[width=\columnwidth,clip,trim=0 3cm 0 0]{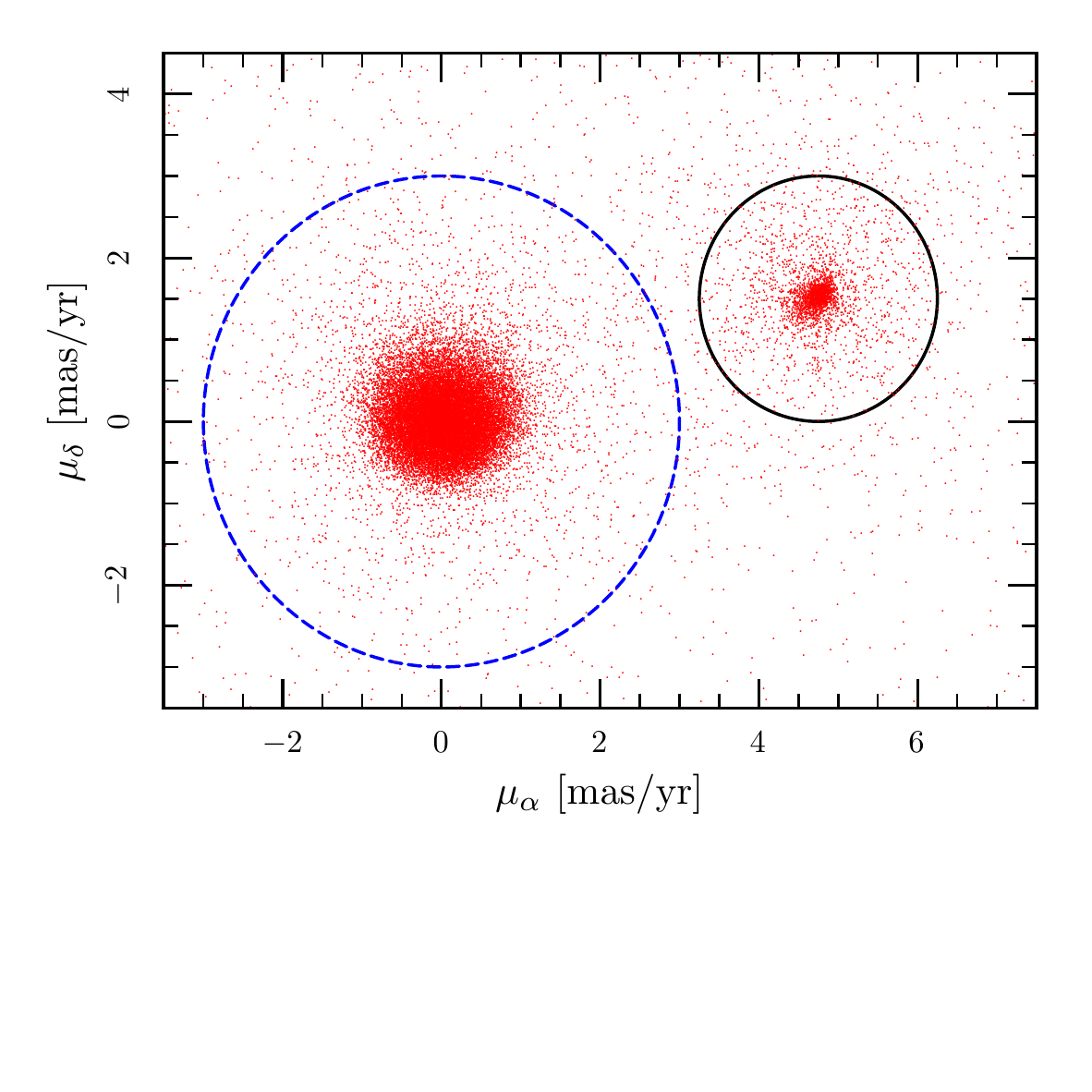}
\caption{Proper motions of the objects identified as stars in the ACS
  field.  Since we used cluster members to define the reference frame,
  the zero point of the vector-point diagram corresponds to the bulk
  motion of the cluster.  Those stars within the dashed blue circle
  comprise the sample to study 47 Tucanae. The stars within the solid
  black circle comprise the sample used study the Small Magellanic
  Cloud.  }
\label{fig:jay_pmall_paper}
\end{figure}

Our team was awarded 121 HST orbits in Cycle 17 to image 47 Tuc (GO-11677, the right field in Figure~\ref{fig:both_field}).  The main science goal was to obtain photometry with the ACS F606W and F814W filters that would go deep enough to study the entirety of the white dwarf cooling sequence. A detailed discussion of the observations and data reduction can be found in \citet{2012AJ....143...11K} and \citet{2008AJ....135.2114A}. The field is centered about 6.7~arcminutes from the core of the cluster and is approximately circular with a diameter of five arcminutes. There were a total of 754 exposures that correspond to thirteen different telescope orientations relative to the cluster, so there is no biased direction in the proper motions. This total includes the new observations and archival observations with both ACS and WFC3 to determine the proper motions. \citet{Rich1347tuc} describe these procedures in further detail. The proper motion of each star is measured relative to the mean motion of other stars in the cluster, so these observations are not directly sensitive to the rotation of the cluster \citep[\cf][] {2003AJ....126..772A}.  Figure~\ref{fig:jay_pmall_paper} depicts the proper motions of all of the stars in the field. Two populations are immediately apparent: the larger group centered on the origin are stars in the cluster, and the smaller group centered at about five milliseconds per year in right ascension are stars in the Small Magellanic Cloud (SMC) that lies in the background of the cluster. We use stars whose proper motion is within three milliarcseconds per year of the mean of the cluster as our cluster sample and stars within 1.5~mas/yr of the mean of the SMC as our SMC sample.

\begin{figure}
\includegraphics[width=\columnwidth]{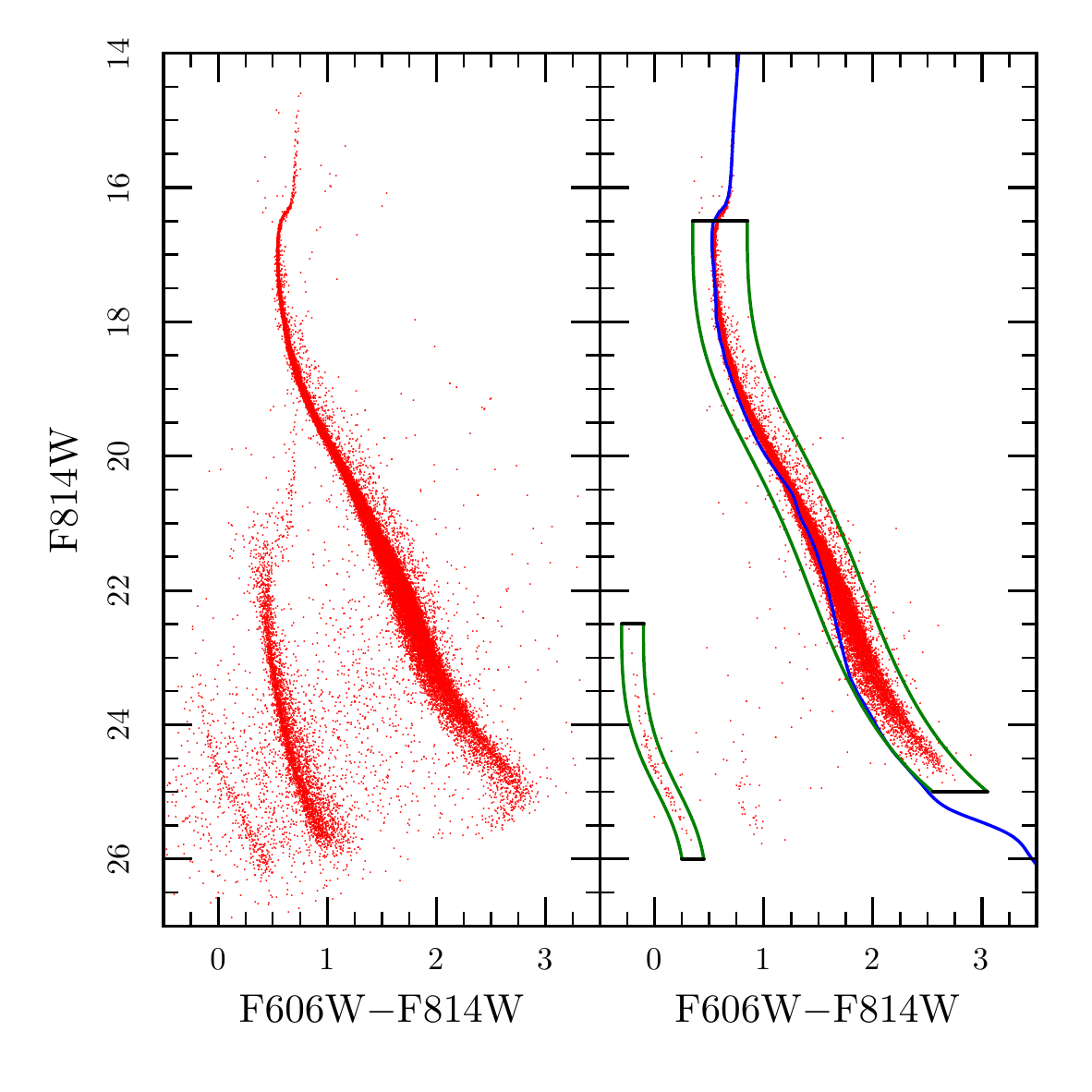}
\caption{The left panel presents the color-magnitude diagram for all
  of the stars in the ACS field.  The right panel shows only those
  stars whose proper motions lie within the dashed blue circle in
  Figure~\ref{fig:jay_pmall_paper}.  Furthermore, the boxes depict
  the white-dwarf and main-sequence subsamples. The blue curve depicts a 11~Gyr-old
  isochrone from MESA.}
\label{fig:jay_cmd}
\end{figure}

Figure~\ref{fig:jay_cmd} depicts the color-magnitude diagram (CMD) for all of the stars in the ACS field in the left-hand panel. The right-hand panel depicts only those stars within the dashed circle in Figure~\ref{fig:jay_pmall_paper} and whose proper-motion error estimate is less than 0.2 mas/yr. This is our proper-motion-cleaned sample of the outer region in 47~Tuc. Furthermore, we divide the sample of 47~Tuc stars into the main-sequence stars in the right-hand box and white-dwarf stars in the left-hand box. We do not examine the proper motions of the stars above the turn-off.  We have superimposed a model isochrone calculated with MESA \citep{2011ApJS..192....3P} and the stellar atmospheres of  \citet{2015A&A...577A..42B} for an age of 11~Gyr for the metallicity of 47~Tuc at a distance of 4.7~kpc  
\citep{2012AJ....143...50W} and the interstellar absorption given in Table~\ref{tab:data}. We use this isochrone to estimate the masses of the stars as a function of F814W. 

\begin{figure}
\includegraphics[width=\columnwidth,clip,trim=0 3cm 0 0]{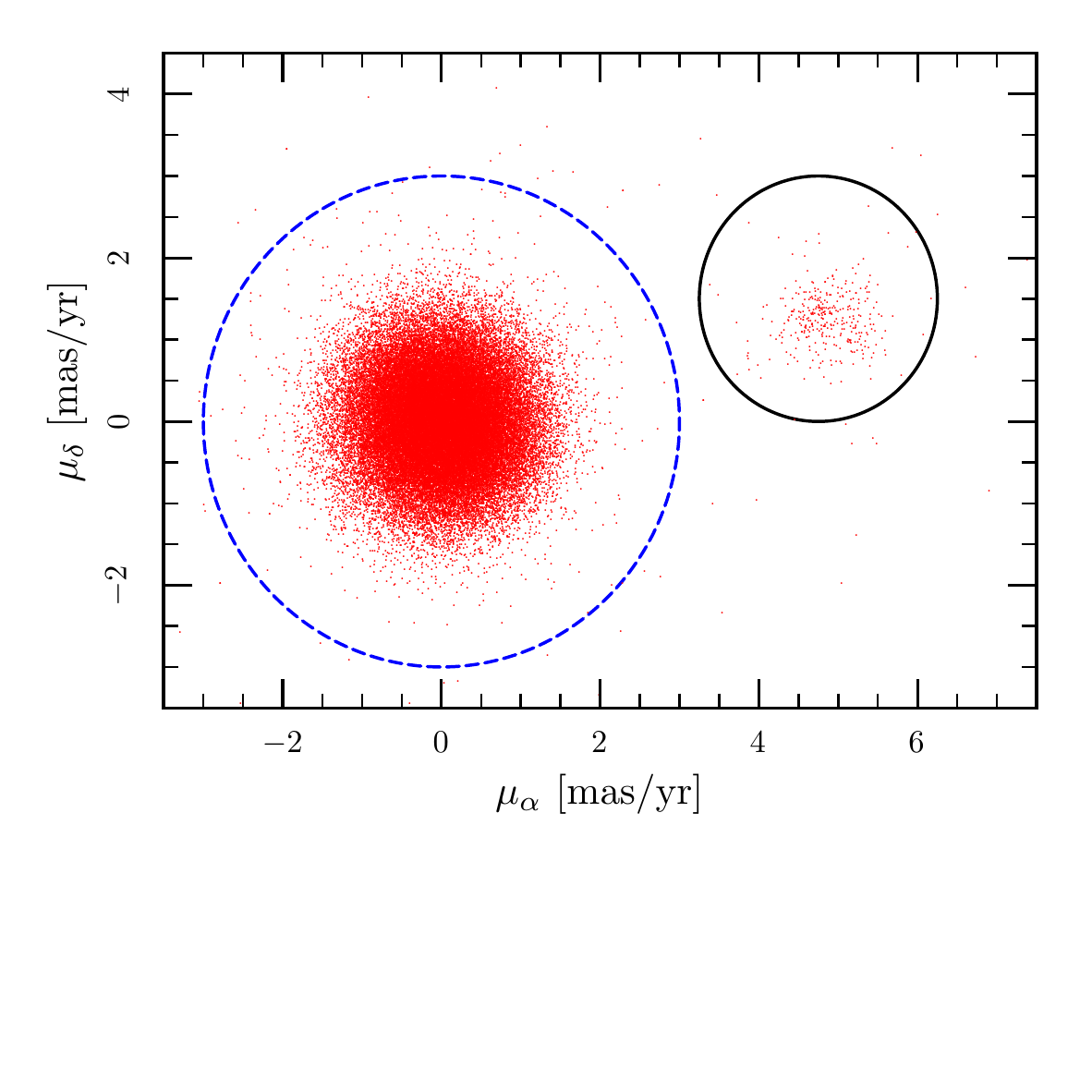}
\caption{Proper motions of the objects identified as stars in the WFC3
  field. Since we used cluster members to define the reference frame,
  the zero point of the vector-point diagram corresponds to the bulk
  motion of the cluster. Those stars within the dashed blue circle
  comprise the sample to study 47 Tucanae.
  The stars within the solid black circle lie within the Small
  Magellanic Cloud.
}
\label{fig:cen_pmall}
\end{figure}

\begin{figure}
\includegraphics[width=\columnwidth]{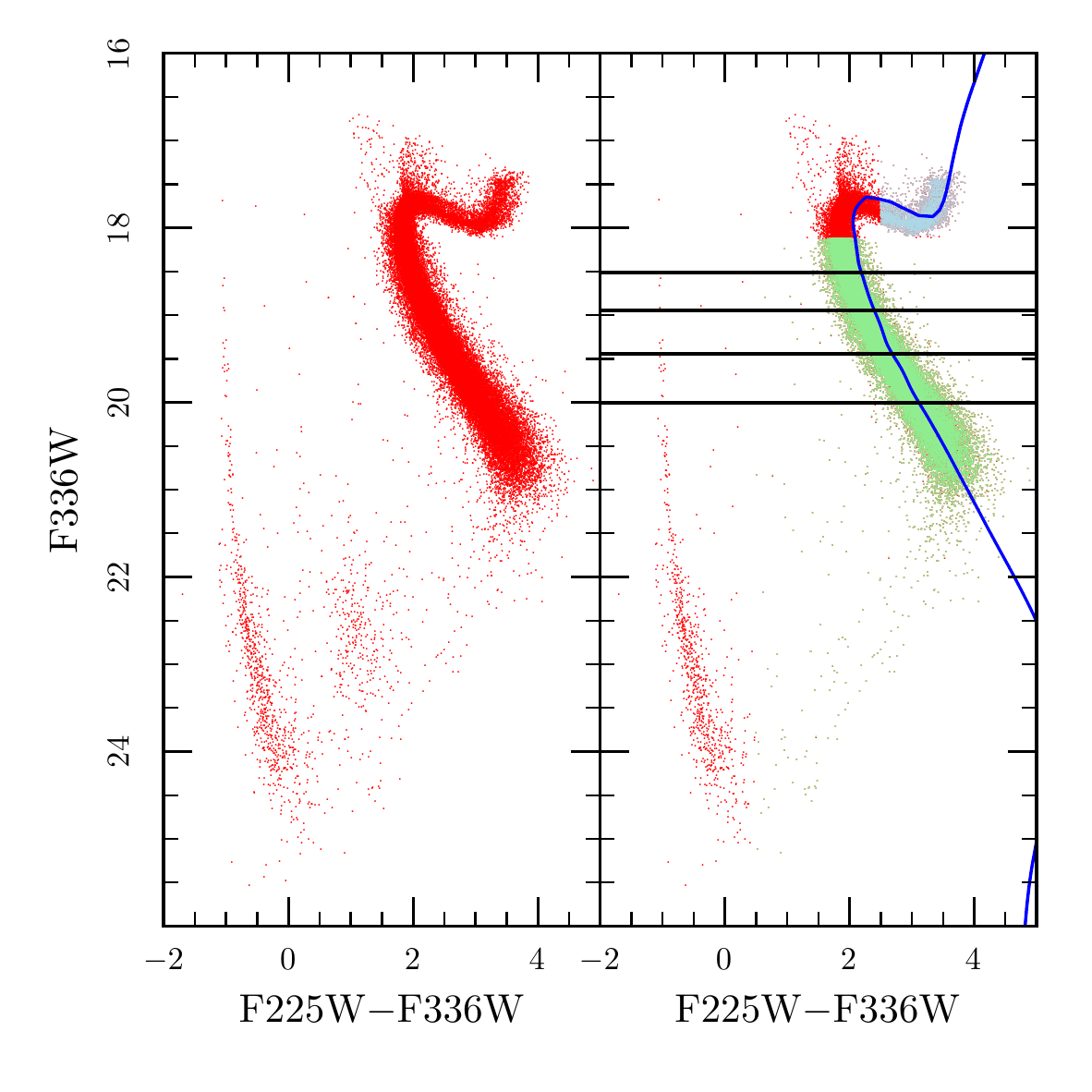}
\caption{The left panel presents the color-magnitude diagram for all
  of the stars in the WFC3 field with proper motions.  The right panel
  shows only those stars whose proper motions lie within the dashed
  blue circle in Figure~\ref{fig:cen_pmall}.  Furthermore, the green
  points depict the main-sequence subsamples divided into magnitude
  bins by the horizontal lines. The light-blue points depict the
  subgiant subsample.  The blue curve depicts a 11~Gyr-old isochrone.}
\label{fig:cen_cmd}
\end{figure}

Subsequently our team was awarded 10 HST orbits in Cycle 20 to image a wide region in the core of 47 Tuc (GO-12971, left field in Figure~\ref{fig:both_field}). The main science goal was to obtain photometry with the WFC3 F225W and F336W filters to characterize a sample of neutrino-cooling white dwarfs in the center of the cluster. We also used archival images of the core of 47 Tuc taken with WFC3 to measure proper motions in the core in a manner similar to the outer field. Figure~\ref{fig:cen_pmall} depicts the observed proper motions in the core of the cluster. Again we see the dramatic distribution of proper motions of 47~Tuc stars centered on the origin along with relatively fewer stars in the SMC. Unlike in Figure~\ref{fig:jay_pmall_paper} the distribution here appears to be circularly symmetric. The left panel of Figure~\ref{fig:cen_cmd}, similar to Figure~\ref{fig:jay_cmd}, depicts the CMD of the stars in the center of the cluster. Here, the stellar density is much higher, so in spite of the relatively bright flux limit, there are many cluster stars both along the white-dwarf cooling track and the main sequence. The contamination from the SMC is relatively mild. Regardless, we remove it through proper-motion cleaning, leaving the CMD in the right panel. Superimposed on the CMD is the same stellar isochrone as in Figure~\ref{fig:jay_cmd} but in the WFC3 bands. In these bands, the fluxes do not increase monotonically with stellar mass, so we restrict our analysis of the main sequence to the stars depicted with green points, where the relationship is monotonic, and we focus on a separate sample of subgiant stars depicted with light-blue points. 

We present the basic properties of the two samples of proper motions in Table~\ref{tab:rmserr}. In the outer field, we divide the sample into the 47~Tuc white dwarfs, the 47~Tuc main sequence stars and the stars of the SMC. Our focus will be the proper motions of the main sequence stars with $18 < \mathrm{F814W} < 24$, whose proper motions are the best determined. We will study the proper motions of these stars as a function of projected radius in \S~\ref{sec:model-47-tucanae} and as a function of F814W in \S~\ref{sec:prop-moti-disp}. We will use the proper motions as a function of projected radius to constrain the theoretical model of the cluster (\S~\ref{sec:model-47-tucanae}). There are many more measured proper motions in the core of the cluster with slightly larger measurement errors than in the outer field. We will study these proper motions as a function of F336W and radius simultaneously. Neither the theoretical modeling nor the statistical techniques that we use here are sufficient to do this rich dataset justice, so we characterize the central proper motions empirically and compare them to the expectations from the dynamical model constrained with the outer proper motions, leaving a more detailed analysis for subsequent work.
 
\begin{table*}
\begin{center}                                                                     \caption{Proper-motion dispersions, root-mean-squared errors, mean errors                        for the various subsamples}
 \label{tab:rmserr}                                                                            
  \begin{tabular}{lrccccccccr}
\hline \hline                                                                                  
      Outer    &        & \multicolumn{2}{c}{Median}  & & & & & Mean & RMS & \multicolumn{1}{c}{$f$} \\
     Sample & Number & F814W & Mass & $\sigma_{p,x,\mu}$ & $\sigma_{p,y,\mu}$ & ${\hat \sigma}(\mu_x)$ &  ${\hat \sigma}(\mu_y)$ & Error & Error & [\%]  \                  
    \\ \hline 
47 Tucanane Samples: \\
% totalmass (~~MS Stars (16.5 $-$ 25))= 5619.24
      ~~MS Stars (16.5 $-$ 25) &  13653 & 21.5 & 0.39  & 0.45 0.45 & 0.43 0.43 & 0.42 & 0.37  & 0.07 & 0.07 &  1.51 \\
% totalmass (~~WD Stars (22.5 $-$ 26))= 14.4187
      ~~WD Stars (22.5 $-$ 26) &    133 & 25.0 & 0.53  & 0.77 0.76 & 0.99 0.97 & 0.71 & 0.88  & 0.16 & 0.17 &  2.77 \\
% totalmass (~~Best MS PMs (18 $-$ 24))= 4856.63
     ~~Best MS PMs (18 $-$ 24) &  12199 & 21.6 & 0.38  & 0.40 0.40 & 0.37 0.37 & 0.40 & 0.35  & 0.06 & 0.07 &  1.31 \\
% totalmass (~~Faint MS Stars (24 $-$ 25))= 71.6967
  ~~Faint MS Stars (24 $-$ 25) &    603 & 24.3 & 0.12  & 1.06 1.05 & 1.11 1.11 & 1.08 & 1.13  & 0.17 & 0.17 &  1.27 \\
SMC Sample: \\
                ~~All MS Stars &   1961 & 23.8 & 0.76  & 0.41 0.38 & 0.37 0.34 & 0.23 & 0.21  & 0.12 & 0.14 & 14.25 \\

\hline \\
\hline \hline                                                                                  
     Core     &        & \multicolumn{2}{c}{Median}  & & & & & Mean & RMS & \multicolumn{1}{c}{$f$} \\
     Sample & Number & F336W & Mass & $\sigma_{p,R,\mu}$ & $\sigma_{p,T,\mu}$ & ${\hat \sigma}_{\mu_R}$ &  ${\hat \sigma}_{\mu_T}$ & Error &  Error & [\%]  \                  
    \\ \hline 
Sub-Giants ($17.2 - 18.1$) &  2932 & 17.9 & 0.89 & 0.57 0.57 & 0.56 0.56 & 0.57 & 0.57  & 0.07 & 0.08 &  0.98 \\
Bin 1 ($18.1 - 18.5$) &   8398 & 18.3 & 0.84 & 0.57 0.57 & 0.57 0.57 & 0.58 & 0.58  & 0.08 & 0.09 &  1.17 \\
Bin 2 ($18.5 - 19.0$) &   8398 & 18.7 & 0.81 & 0.57 0.57 & 0.57 0.56 & 0.58 & 0.58  & 0.08 & 0.09 &  1.24 \\
Bin 3 ($19.0 - 19.4$) &   8398 & 19.2 & 0.78 & 0.59 0.59 & 0.57 0.57 & 0.59 & 0.58  & 0.08 & 0.10 &  1.29 \\
Bin 4 ($19.4 - 20.0$) &   8398 & 19.7 & 0.74 & 0.58 0.58 & 0.58 0.58 & 0.58 & 0.58  & 0.09 & 0.10 &  1.50 \\
Bin 5 ($20.0 - 24.6$) &   8399 & 20.4 & 0.70 & 0.60 0.60 & 0.59 0.59 & 0.60 & 0.58  & 0.10 & 0.11 &  1.72 
  \end{tabular}
\end{center}                                                                                   

\smallskip

The values in the columns labeled $\sigma_p$ contain both the directly calculated population standard deviation and a value determined using the maximum likelihood technique of \citet{2006AJ....131.2114W}. The quantity $f=1 - {\hat \sigma}_{n,\mathrm{true}}/{\hat   \sigma}_{n,\mathrm{obs}}$ gives the relative decrease in the value of the velocity dispersion after correcting for the uncertainty in the observed proper motions.  The outer sample only includes the half of the stars with the best measured proper motions as a function of apparent magnitude as shown by the red points in Figure~\ref{fig:pmrmag}. The core samples have an proper-motion-error cutoff of 0.4 mas/yr.  The masses of main-sequence stars are given in solar masses using the models described in the text and assuming a distance of 4.7~kpc.  The masses for the white dwarfs come from the spectroscopic measurements of \citet{2008ApJ...676..594K}. The magnitude ranges are in F814W for the outer sample and in F336W for the core sample.
\end{table*}

% \subsection{The ACS Field}
% \label{sec:acs-field}

% \begin{figure}
% \includegraphics[width=\columnwidth]{field_xy}
% \caption{Hello}
% \label{fig:field}
% \end{figure}

%\begin{figure}
%\includegraphics[width=\columnwidth]{windowfunk}
%\caption{Window function}
%\label{fig:windowfunk}
%\end{figure}

\section{Theoretical Model}
\label{sec:modelling-47-tucanae}

To obtain estimates of the total mass of the cluster and the escape velocities of individual stars, we will rely on a theoretical model of the cluster. We will use the model as a method of deprojecting our velocity and surface density measurements to constrain the dynamical properties of the cluster. This section outlines the model.
% and the statistical measures that we will used to characterize both the model
% and the data. 
An appendix derives several useful results from the model that further
our interpretation of the data.

To model the stars in the globular cluster, we will assume that the phase-space density is given by a lowered Maxwellian distribution \citep{1963MNRAS.126..499M,1966AJ.....71...64K} with modifications to allow for velocity anisotropy and rotation \citep{1987AJ.....93.1106L},
\begin{eqnarray}
f&=&\frac{\d N}{\d^3 x \d^3 v} \label{eq:1}
\\
&=&\rho_1 \left(2\pi \sigma^2\right)^{-3/2} e^{\omega J_z/\sigma^2}
\! e^{-J^2/(2\sigma^2 r_a^2)}\!
 \left (e^{\epsilon/\sigma^2}-1\right )~
\label{eq:2}
\end{eqnarray}
where $\rho_1$ is the normalization, $\epsilon=\Psi - \frac{1}{2}v^2$ is the negative of the energy of a star per unit mass. The energy is less than zero because the stars are bound; therefore, $\epsilon$ is positive. The parameter $\sigma$ has dimensions of velocity and sets the velocity scale for a particular group of stars in the cluster, $J^2$ is the square of the total angular momentum per unit mass, $J_z$ is the $z-$component of the angular momentum per unit mass, and $\Psi$ is the negative of the gravitational potential. For simplicity, in the modeling we will assume that $\sigma$ is constant for all of the stars that we have observed and measure the deviations from this assumption. The \citet{1987AJ.....93.1106L} model assumes that both the total angular momentum and the $z-$component of the angular momentum are integrals of the motion. This is a reasonable approximation for small rotation rates, but as the globular cluster becomes more and more flattened by the rotation, this approximation is less and less appropriate.

The definition of the distribution function includes the condition of non-negativity, so if the value of $\epsilon$ becomes positive, the value of $f$ is assumed to vanish. If we let the zeropoint of the energy be the value of the gravitational potential at the tidal radius, we can define the escape velocity from any point in the cluster as $v_e^2/2=\Psi({\bf r})$ where $\Psi({\bf r})$ is the negative of the gravitational potential.  This vanishes at the edge of the cluster (the tidal surface).  To calculate the density and moments of the Lupton-Gunn distribution function requires numerical quadrature. In the appendix, we derive several useful closed-form expressions using the less general Michie-King model or anisotropic lowered isothermal profile \citep{1963MNRAS.125..127M,1966AJ.....71...64K} where $\omega=0$.

The velocity distribution is approximately isotropic within the radius $r_a$ and becomes dominated by radial orbits for $r>r_a$. Finally, the parameter $\omega$ is the angular velocity that characterizes the rotational motion. We can also define a typical radius for the rotational component of the motion $r_b=\sigma/2\omega$, yielding
\begin{eqnarray}
f&=&\rho_1 \left(2\pi \sigma^2\right)^{-3/2} 
 e^{v_\phi R/(2 \sigma r_b)} \times \nonumber \\
 & &  e^{-(v_\phi^2+v_\theta^2) r^2/(2\sigma^2 r_a^2)}\!
 \left (e^{\epsilon/\sigma^2}-1\right ) \label{eq:2b} \\
 &=& \rho_1 \left(2\pi \sigma^2\right)^{-3/2} \Biggr \{ 
 e^{-[(v_\phi-\bar v)^2+v_\theta^2]/(2\sigma_t^2)} e^{-v_r^2/(2\sigma^2)}
 \times \nonumber \\
 & & 
  e^{\Phi/(2\sigma^2)} - e^{v_\phi R/(2 \sigma r_b)} e^{-(v_\phi^2+v_\theta^2) r^2/(2\sigma^2 r_a^2)} \Biggr \}
\label{eq:2c}
\end{eqnarray}
We can examine Eq.~\ref{eq:2c} in the limit where $\Phi \gg\ \sigma^2$ so that the first term in the braces dominates and the distribution is still approximately Maxwellian.  We see that the stellar distribution will exhibit two different velocity dispersions with the smaller dispersion in the tangential direction where
\begin{equation}
\sigma_t = \sigma \frac{r_a}{\sqrt{r^2+r_a^2}}
\label{eq:sigmat}
\end{equation}
and the azimuthal velocity ($v_\phi$) is centered around
\begin{equation}
\bar v = \frac{1}{2} \frac{R}{r_b} \frac{\sigma_t^2}{\sigma}.
\label{eq:meant}
\end{equation}
Both the anisotropy of the velocity ellipsoid and the mean rotational velocity increase with radius.   The second term in the braces causes the distribution of velocities to be normal no longer.  We will examine the third moment of the velocity distribution, the skewness, which appears because of the rotation; in fact because our measured proper motions use the mean motion of the cluster as a reference it is through this skewness that measurements in a single field can be sensitive to the rotation.  However, when there is both anisotropy and rotation, the skewness of the distribution (Eq.~\ref{eq:2c}) is limited to values of about one tenth.

We will generally use bootstrapping of the stellar samples to estimate the errors in our statistical quantities \citep{Heyl116397dyn,1993stp..book.....L} and the robust $\hat \sigma$ estimator for the velocity dispersion \citep{Heyl116397dyn,Rous93}.  Because we will restrict our samples of proper motions to those with the smallest error estimates, the typical proper motion error of the stars is much less than the proper motion dispersion.  In this limit, the maximum likelihood estimator for the velocity dispersion introduced by \citet{2006AJ....131.2114W} reduces to the difference in quadrature of the standard deviation of observed proper motions and the root-mean-squared of the proper motion errors for the sample.  We denote this difference the population standard deviation, $\sigma_p$.  In Table~\ref{tab:rmserr} we give both the maximum likelihood value and the robust estimator for the proper motion dispersions.  The maximum likelihood estimator is very sensitive to outliers whereas the robust estimator can tolerate up to a quarter of data points as outliers.  For the large samples, the \citet{2006AJ....131.2114W} estimator and the population standard deviation yield similar values as shown in the columns labeled $\sigma_p$ in Table~\ref{tab:rmserr}.  This is not surprising because our typical proper motion errors are one-tenth the population standard deviation.  The largest differences occur in the SMC sample where the RMS proper motion error is about one-third of the population standard deviation. Because we want to be careful about outliers and we have strong error cuts, we use the robust estimator $\hat \sigma$ throughout \citep{Rous93}. 

\section{Results}
\label{sec:results}

\subsection{A Model for 47~Tucanae}
\label{sec:model-47-tucanae}

\begin{figure}
\includegraphics[width=\columnwidth]{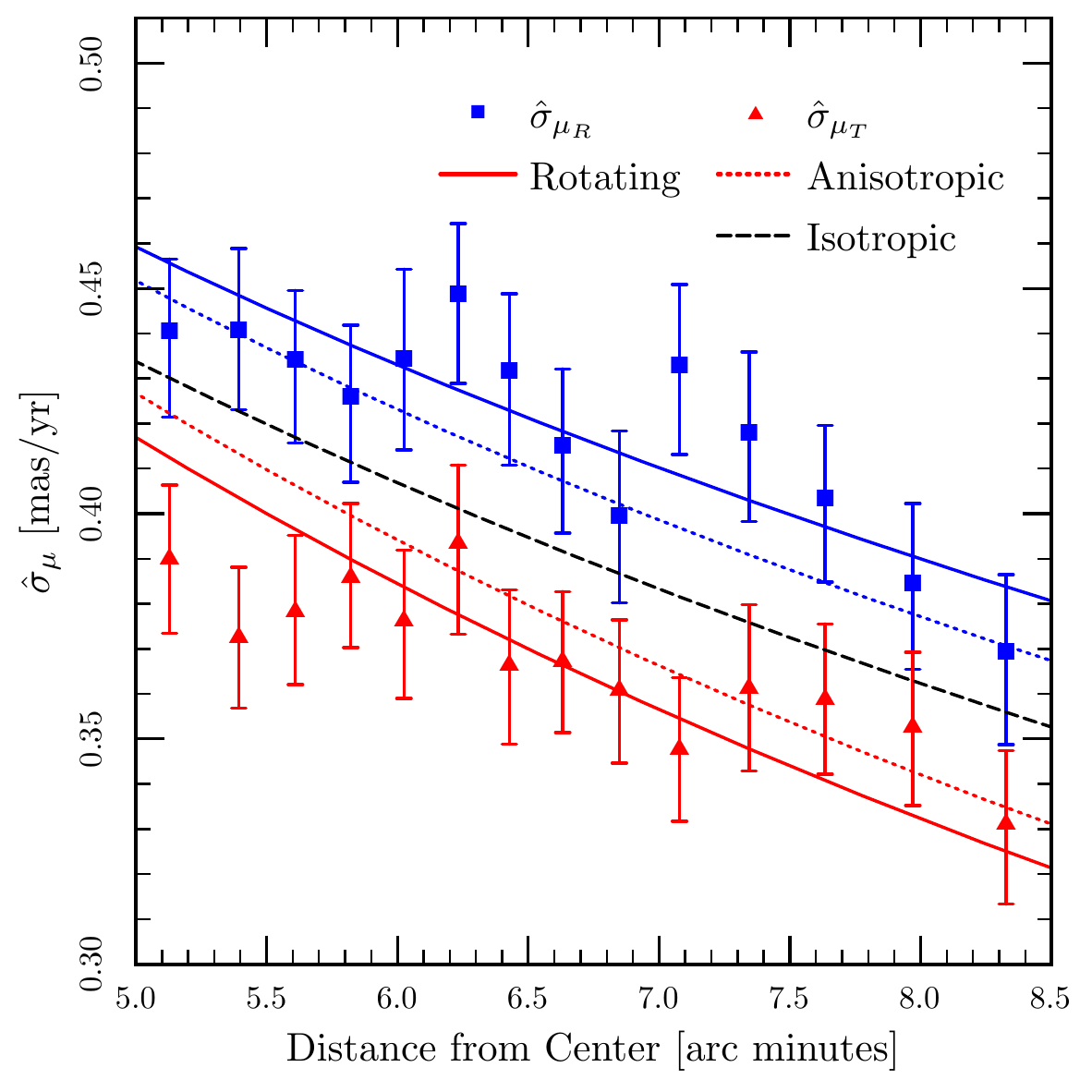}
  \includegraphics[width=\columnwidth]{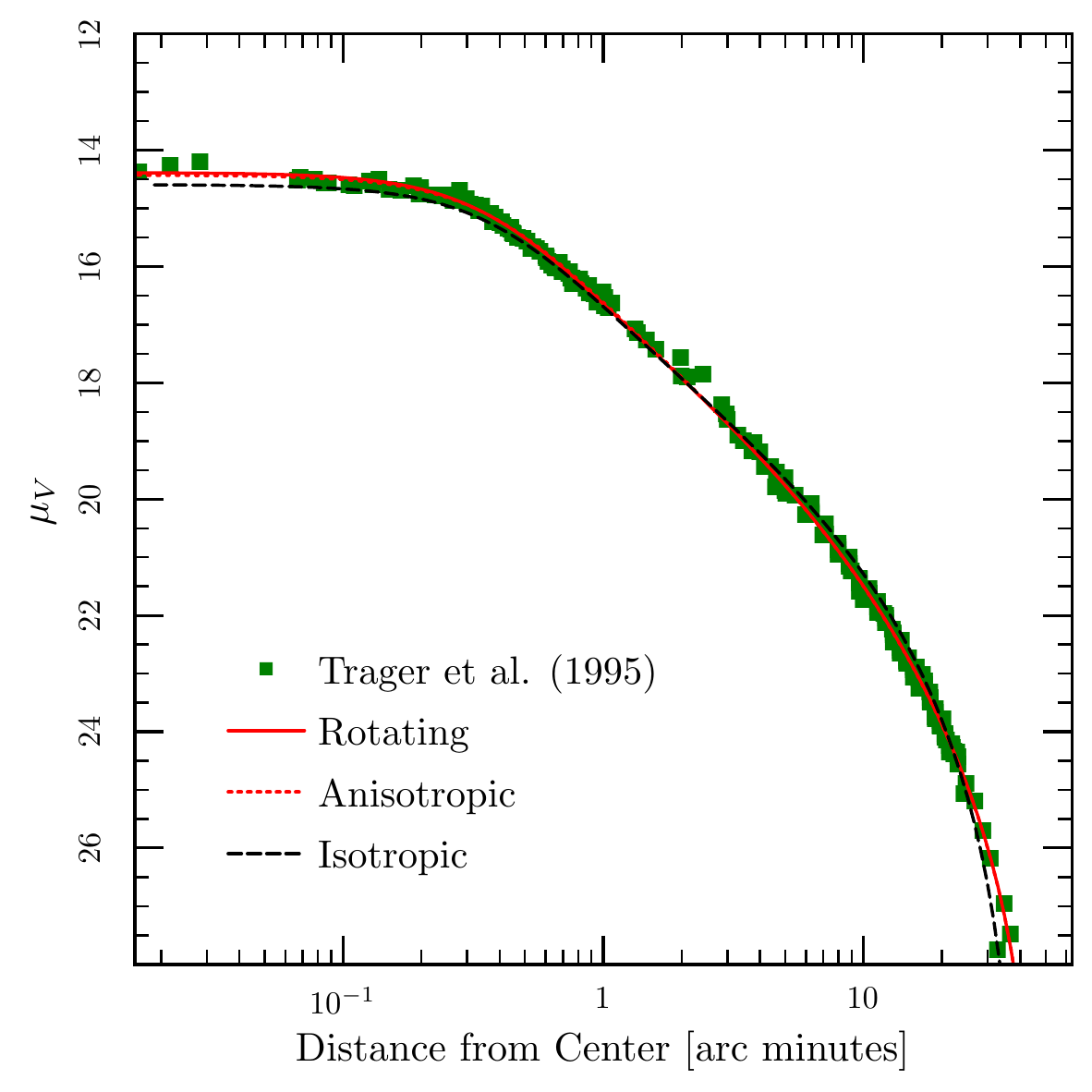}
  \caption{ The model fits to the observed velocity dispersion (upper) surface brightness profile (lower). The solid lines trace the rotating anisotropic model, the dotted lines trace the non-rotating anisotropic model, and the dashed line traces the isotropic model.  Upper panel: The value of ${\hat \sigma}$ as a function of projected radius   calculated in this paper.  Lower panel: The surface brightness profile measured by \citet{1995AJ....109..218T} in black.  }
\label{fig:best_models}
\end{figure}

With a sample of about 12,000 main-sequence stars with $18 < \mathrm{F814W} < 24$ in our outer field, we sort the stars in projected radial distance from the center of the cluster and determine the dispersion as a function of radius as shown in the upper panel Figure~\ref{fig:best_models}. We only use the stars whose proper motion errors are smaller than the median proper motion error at the star's observed magnitude.  That is, we use the best measured proper motions.  Each subsample consists of about 500 stars. The confidence intervals are determined by bootstrap resampling, and the root-mean-square error in proper motion is subtracted from ${\hat \sigma}$ as discussed in \citet{Heyl116397dyn}. Especially at the faint end, our sample is not complete.  However, we do not expect the completeness rate to be a function of the proper motion of the stars.  Because we are focus wholly on the proper-motion distribution, we do not need to correct for completeness.

We calculate a grid of globular cluster models with $\Psi(0)/\sigma^2$ ranging from 7.5 to 9.5 and $(r_a)^{-1}$ with spacing of $0.2$ and $(r_b)^{-1}$ from 0 to $(10~\mathrm{arcminutes})^{-1}$ with a spacing of about $(1~\mathrm{arcminute})^{-1}$.  We calculate the value of $\chi^2$ for the various models using the surface brightness profile of 47~Tucanae from \citet{1995AJ....109..218T} and the velocity dispersions in our ACS field as depicted in Figure~\ref{fig:best_models}.   We use the prescription outlined by \citet{2005ApJS..161..304M} to estimate the uncertainties in the \citet{1995AJ....109..218T} surface brightness measurements and that the fractional uncertainty in the proper-motion dispersion is 0.03. The value of $\chi^2$ is much more strongly dependent on the value of $\Psi(0)/\sigma^2$ than the other two parameters, so we calculate additional models better to determine the value of this parameter.  We also must fit for the overall scale on the cluster in angular size, proper motion and mass-to-light ratio. We fit isotropic models (the standard King model), anisotropic models (Michie-King models) and rotating models (Lupton-Gunn models).  For the rotating models, we make the simplifying assumption that we are looking down the rotation axis of the cluster.  \citet{2017ApJ...844..167B} argue that the inclination of the rotation axis to the line of sight is about 30~degrees.  However, within their models the proper motions are better fit by smaller inclination angles and the line-of-sight velocities by larger ones. We will compare the results from the model with the various observations of our outer sample in the subsections that follow.  All three classes of models can fit the surface brightness profile, but only the anisotropic and rotating anisotropic models can fit the proper motions which are anisotropic as apparent in Figure~\ref{fig:jay_pmall_paper} and~\ref{fig:best_models}. The rotating models yield the best fit to the velocity dispersions, but the non-rotating anisotropic models also yield an acceptable fit.  Table~\ref{tab:model_data} presents the model parameters for the best-fitting models of the three classes.  All the models yields a similar total mass for the cluster of about $1.4 \times 10^6 d^3_{4.7}~\mathrm{M}_\odot$.

To determine how the assumed inclination angle affects the estimate of the mass of the cluster,  we also calculated a suite of models with an inclination angle of 25 degrees.  This value appears to be the best favored by the \citet{2017ApJ...844..167B} proper-motion data (see the left panel of their Figure~6).  We assume that the projection of the rotation axis onto the sky points along the direction connecting the center of the cluster with our outer field.   In reality the rotation axis is thought to point about 45 degrees south of this direction \citep[see Figure~4][]{2017ApJ...844..167B}.   The results for this model are given in the final row of Table~\ref{tab:model_data}.   The value of the $\sigma$ parameter is slightly lower and the model is also slightly more compact, yielding a slightly lower mass estimate of $1.3 \times 10^6 d^3_{4.7}~\mathrm{M}_\odot$.
% replaced on 3 October 2017 
%\begin{table}
%\begin{center}
%\caption{Model Parameters for 47 Tucanae}
%\label{tab:model_data}
%\begin{tabular}{lcccccc}
%\hline
%Model & $\Psi(0)/\sigma^2$ & $\mu_e(0)$ & $\sigma$ & $r_a$ & $r_b$ & Mass \\
%\hline
%Isotropic   & 8.65              & 2.71                 & 0.65 &     &  & 1.1 \\
%Anisotropic & 8.32              & 2.95                 & 0.72 & 7.8 &  & 1.1 \\
%Rotating    & 8.3\phantom{2}  & 2.6\hspace{1.35mm}   & 0.65 & 12.\hspace{2.5mm} & %12.\hspace{3.5mm} & 1.6 \\
%% Rotating 2    & 8.3\phantom{2}  & 2.44                 & 0.60 & 11.\hspace{2.5mm} &  9.2 & 1.3 \\
%\hline
%\end{tabular}
%\end{center}
%The velocities have units of mas~yr$^{-1}$, the radii have units of arcminutes and mass in %units of $10^6 d_{4.7}^3 \mathrm{M}_\odot$.  The quantity $r_b$ is given by the expression %$\sigma/\omega$ and $\mu_e(0)$ is the value of the escape proper motion from the center of %the cluster. Smaller values of $r_a$ and $r_b$ indicate a larger contribution from
%anisotropy and rotation respectively.
%\end{table}
\begin{table}
\begin{center}
\caption{Model Parameters for 47 Tucanae}
\label{tab:model_data}
\begin{tabular}{lcccccc}
\hline
Model & $\Psi(0)/\sigma^2$ & $\mu_e(0)$ & $\sigma$ & $r_a$ & $r_b$ & $\chi^2/n$ \\
\hline
Isotropic   & 8.6                & 2.74                 & 0.66 &     &     & 5.24\\
Anisotropic & 8.4                & 2.95                 & 0.72 & 6.1 &     & 1.26\\
Rotating    & 8.6                & 2.90                 & 0.70 & 5.6 & 6.5 & 1.19 \\
Inclined Rotating    & 8.6                & 2.87        & 0.69 & 5.6 & 6.5 & 1.20 \\
% Isotropic   & 8.8              & 2.77                 & 0.66 &     &     & 6.26 \\
% Anisotropic & 8.4              & 2.95                 & 0.72 & 6.1 &     & 1.65\\
% Rotating    & 8.6              & 2.48                 & 0.60 & 5.5 & 6.4 & 1.60 \\
% Rotating 2    & 8.3\phantom{2}  & 2.44                 & 0.60 & 11.\hspace{2.5mm} &  9.2 & 1.3 \\
\hline
\end{tabular}
\end{center}
The velocities have units of mas~yr$^{-1}$ and the radii have units of arcminutes.  All of the models have a total mass of approximately $1.4\times 10^6 d_{4.7}^3 \mathrm{M}_\odot$.  The quantity $r_b$ is given by the expression $\sigma/2\omega$ and $\mu_e(0)$ is the value of the escape proper motion from the center of the cluster. Smaller values of $r_a$ and $r_b$ indicate a larger contribution from anisotropy and rotation respectively.  $n$ is the number of degrees of freedom in the fit (about 230).
\end{table}

\subsection{Mass Segregation in Proper-Motion}
\label{sec:prop-moti-disp}

The models discussed in \S~\ref{sec:model-47-tucanae} use the same value of $\sigma$ for all of the stars; therefore, the velocity dispersion for all of the stellar populations necessarily have the same value.  Here we will examine the proper-motion dispersion as a function of magnitude in the ACS field; we sort the sample of about 14,000 main-sequence stars of 47 Tucanae by apparent magnitude into subsamples of 1,000 and the nearly 2,000 main-sequence stars of the SMC into samples of 400 stars.  We use the proper motion of the SMC stars to determine whether our estimated uncertainties in the measured proper motions are reasonable.  Because these stars range in mass from about 0.6 to 1 solar mass and furthermore the SMC is not expected to be dynamically relaxed, we do not expect the proper motion dispersion of these stars to increase dramatically toward fainter magnitudes.  We also focus on the faintest stars along the main sequence of 47 Tuc with $\textrm{F814W}>24$.  These number 603 and are divided into samples of 200 stars.  We calculate the value of ${\hat \sigma}$ for each subsample in the projected radial (${\hat \sigma}_{\mu_R}$) and tangential directions (${\hat \sigma}_{\mu_T}$).    The root-mean-squared error in the proper motion is scaled and subtracted from each estimate of the dispersion (and its confidence interval) in quadrature \citep[see \S3.3 of][for more details]{Heyl116397dyn}.
\begin{figure}
\includegraphics[width=\columnwidth]{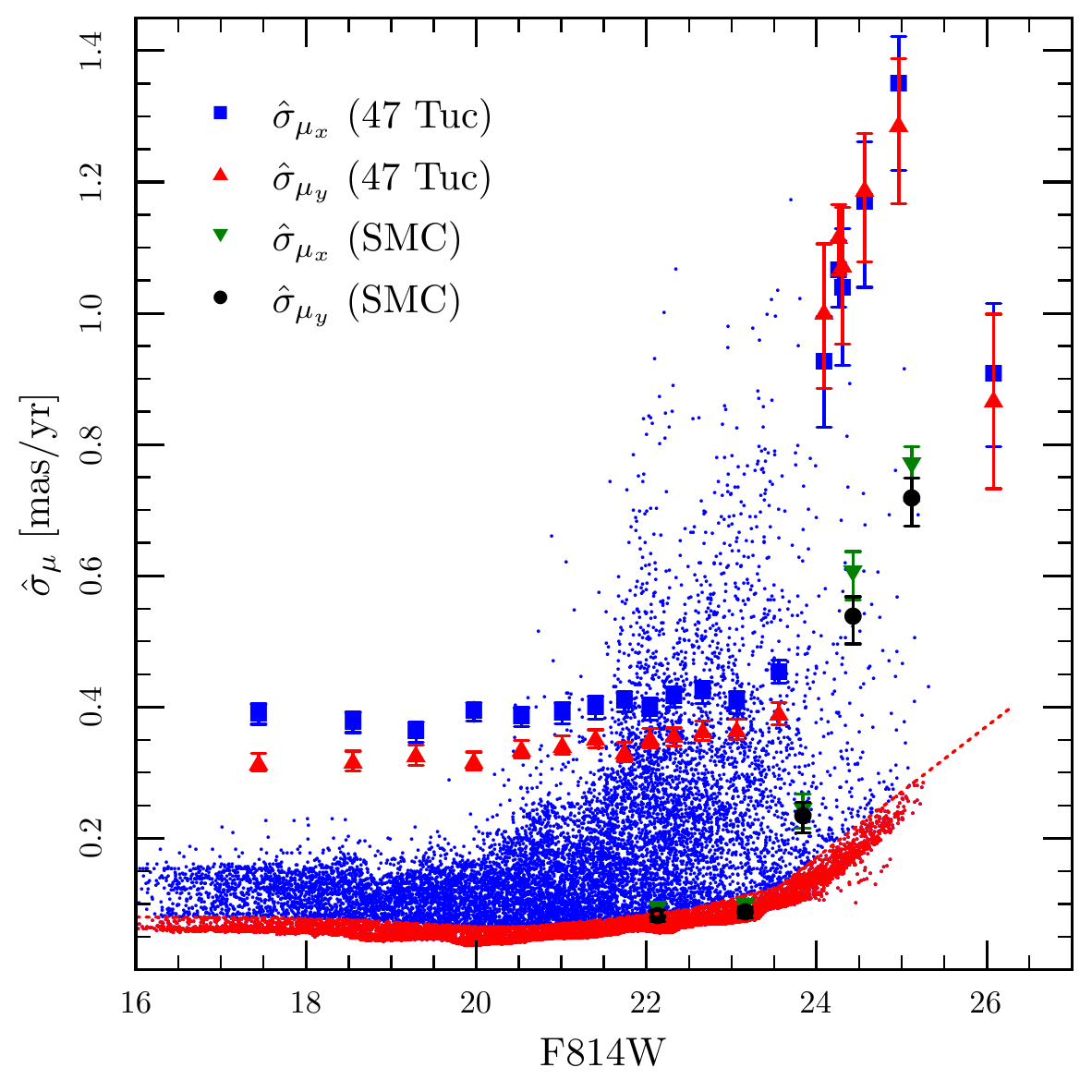}
\includegraphics[width=\columnwidth]{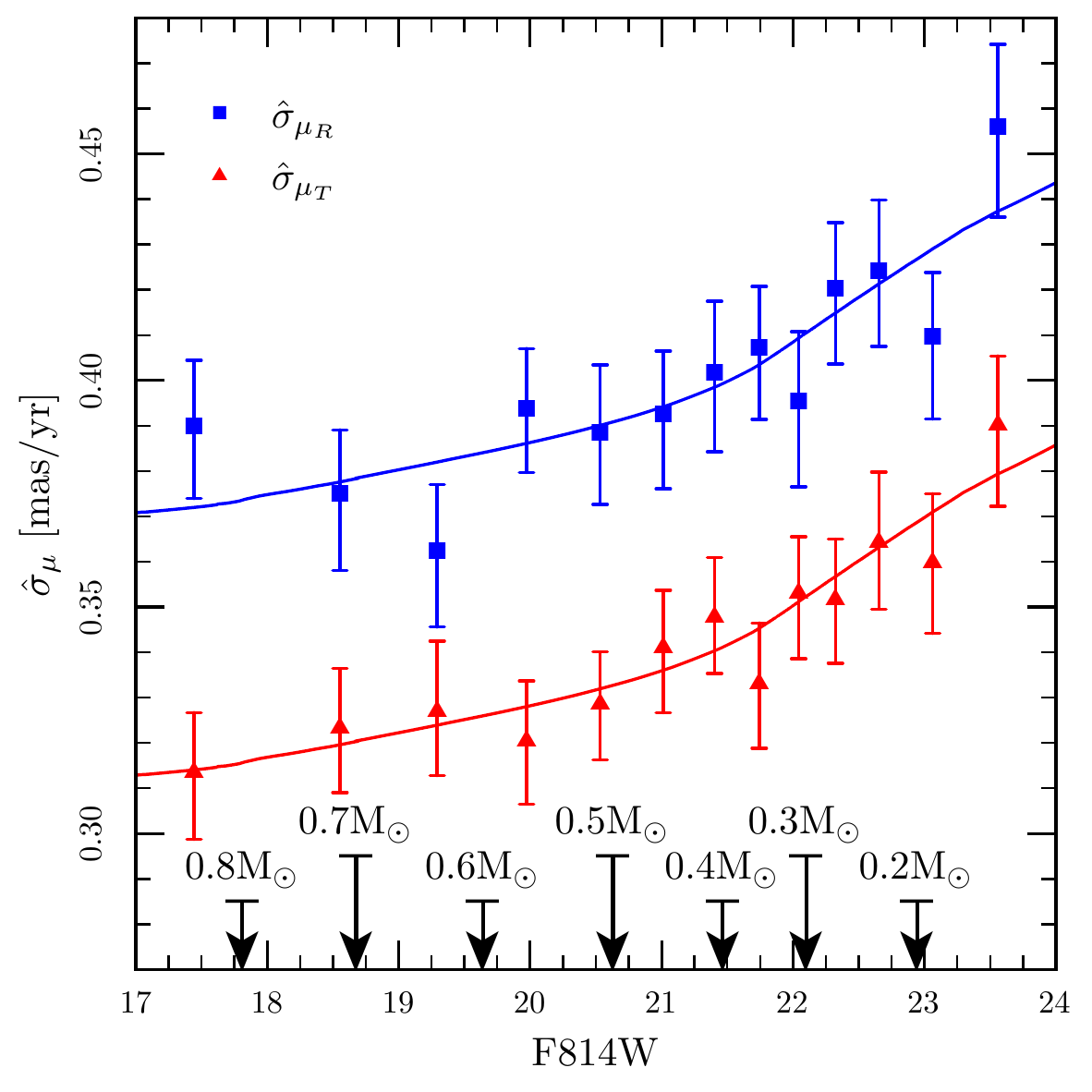}
\caption{Upper panel: The value of ${\hat \sigma}$ as a function of apparent magnitude for stars in 47~Tuc and the SMC.  The blue points depict the proper-motion error estimates of all of the stars, and the red points depict the uncertainties for the stars included in the sample.  There is a large increase in the proper-motion dispersion of the 47 Tuc stars fainter than F814W of 24.  Lower panel: The value of ${\hat \sigma}$ and mass (from the MESA models) as a function of apparent magnitude for 47~Tuc stars brighter than $\mathrm{F814W}=24$.  The arrows in the lower plot indicate the apparent magnitude of stars from 0.8 to 0.2 solar masses.  The solid curves give the power-law fits for the dispersion Eqs.~\ref{eq:3} and~\ref{eq:4}.}
\label{fig:pmrmag}
\end{figure}

The upper panel of Figure~\ref{fig:pmrmag} depicts the proper motion dispersions along the direction of right ascension ($\hat \sigma_{\mu_x}$ --- this is the motion across the sky not the change in the coordinate, \ie it has been corrected with the cosine of the declination) and declination ($\hat \sigma_{\mu_y}$) for both the stars in the SMC and in 47 Tuc --- these are approximately the radial and tangential directions.  The proper motion dispersion of the bright main sequence stars is typically about 0.4~mas/yr and exhibits significant anisotropy.  The proper motion dispersion of the stars of the SMC is typically much smaller about 0.1~mas/yr.  Fainter than $\mathrm{F814W}\approx 24$ the dispersion of both the SMC and 47 Tuc stars increases dramatically.  This indicates that we may have underestimated the proper motion errors of these faint stars because we expect the proper-motion dispersion of the SMC stars to be constant with stellar mass as it is not a relaxed population.
% To test this hypothesis we assume that the increase in the observed dispersion of the stars in the SMC may be accounted by an unmodeled increase in the proper motion errors.  If we assume that the actual proper motion dispersion of the 47~Tuc stars remains constant fainter than $\mathrm{F814W}\approx 24$ and the increase is due to unmodeled errors, the resulting observed proper motion dispersion would follow the upper blue and red curves.  These curves do not reach the observed dispersion of the 47~Tuc stars, indicating that the dispersion of these faint stars may be 50\% larger than the brighter ones. 
Because of this uncertainty in the quality of the proper-motion measurements of these stars, we have excluded the stars with $\mathrm{F814W}>24$ from our analysis.
 
The lower panel of Figure~\ref{fig:pmrmag} shows the proper motion dispersion in the projected radial and tangential directions relative to the center of the cluster.  The figure shows that the proper-motion dispersion is nearly constant, increasing slowly with fainter magnitudes along the main sequence.  We fit the observed proper motion dispersions with a power-law function of mass to characterize the trends.  For the radial proper motion dispersion we find
\begin{equation}
  \hat{\sigma}_{\mu_R} \approx 0.37 \left (\frac{M}{\mathrm{M}_\odot} \right )^{-0.096}
  \label{eq:3}
\end{equation}
and for the tangential dispersion
\begin{equation}
  \hat{\sigma}_{\mu_T} \approx 0.31 \left (\frac{M}{\mathrm{M}_\odot} \right )^{-0.11}.
\label{eq:4}
\end{equation}
The results of these fits are depicted in Figure~\ref{fig:pmrmag} as
solid curves.  We can also fit both dispersions with a common exponent
of $-0.10$ and in this case the normalizations are $0.36$ and $0.31$
respectively.

% (array([ 0.36566773, -0.09604334]), 1) rad
% (array([ 0.30784052, -0.11209739]), 1) tan
% (array([ 0.36317493,  0.31072285, -0.10284707]), 1) combo with same exponent

\subsection{Proper-Motion Distribution}
\label{sec:prop-moti-distr}

\begin{figure}
\includegraphics[width=\columnwidth]{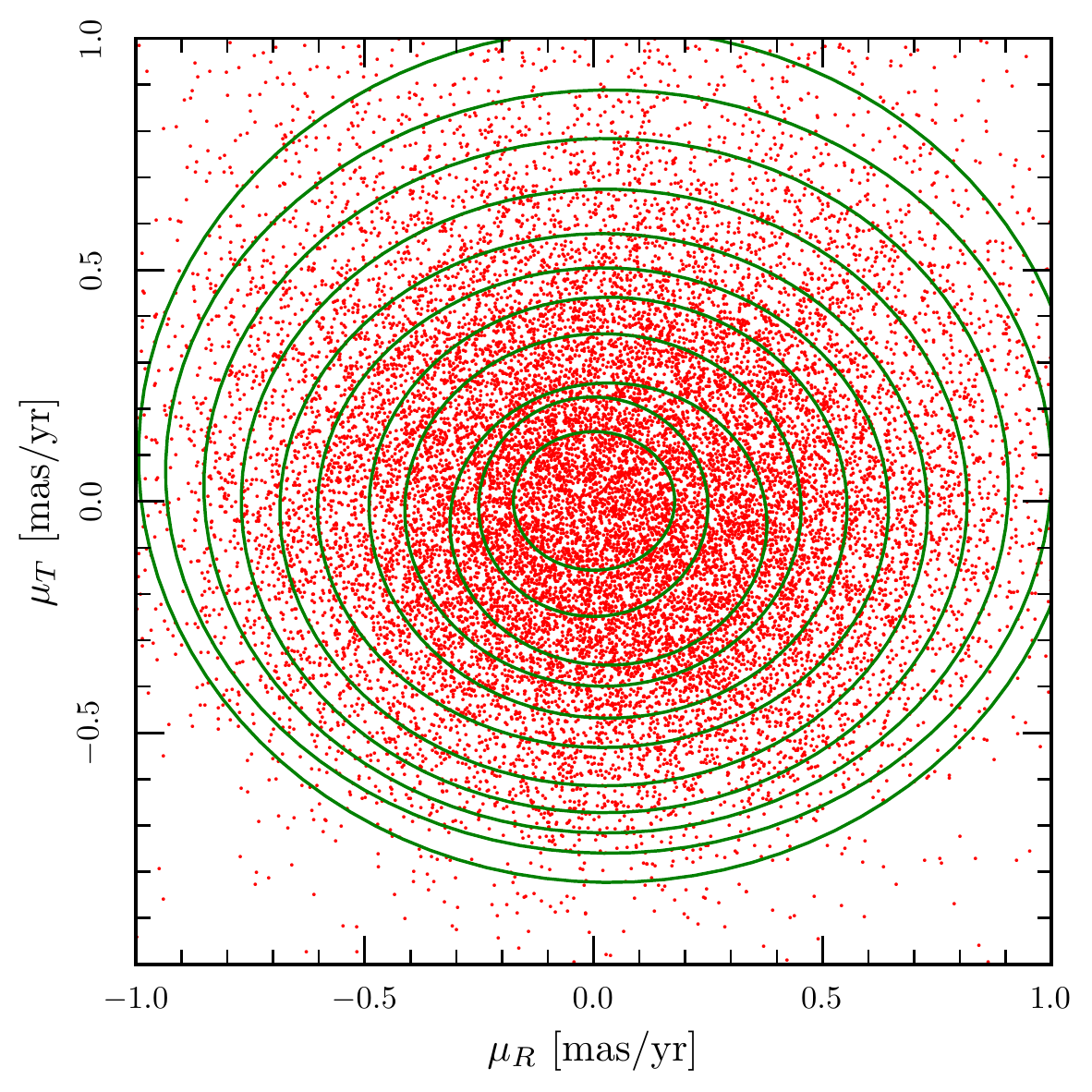}
\caption{The best-fitting ellipsoidal contours of equal phase-space
  density as a function of the radial and tangential proper motion}
\label{fig:drawcontourr}
\end{figure}

\begin{figure}
\includegraphics[width=\columnwidth]{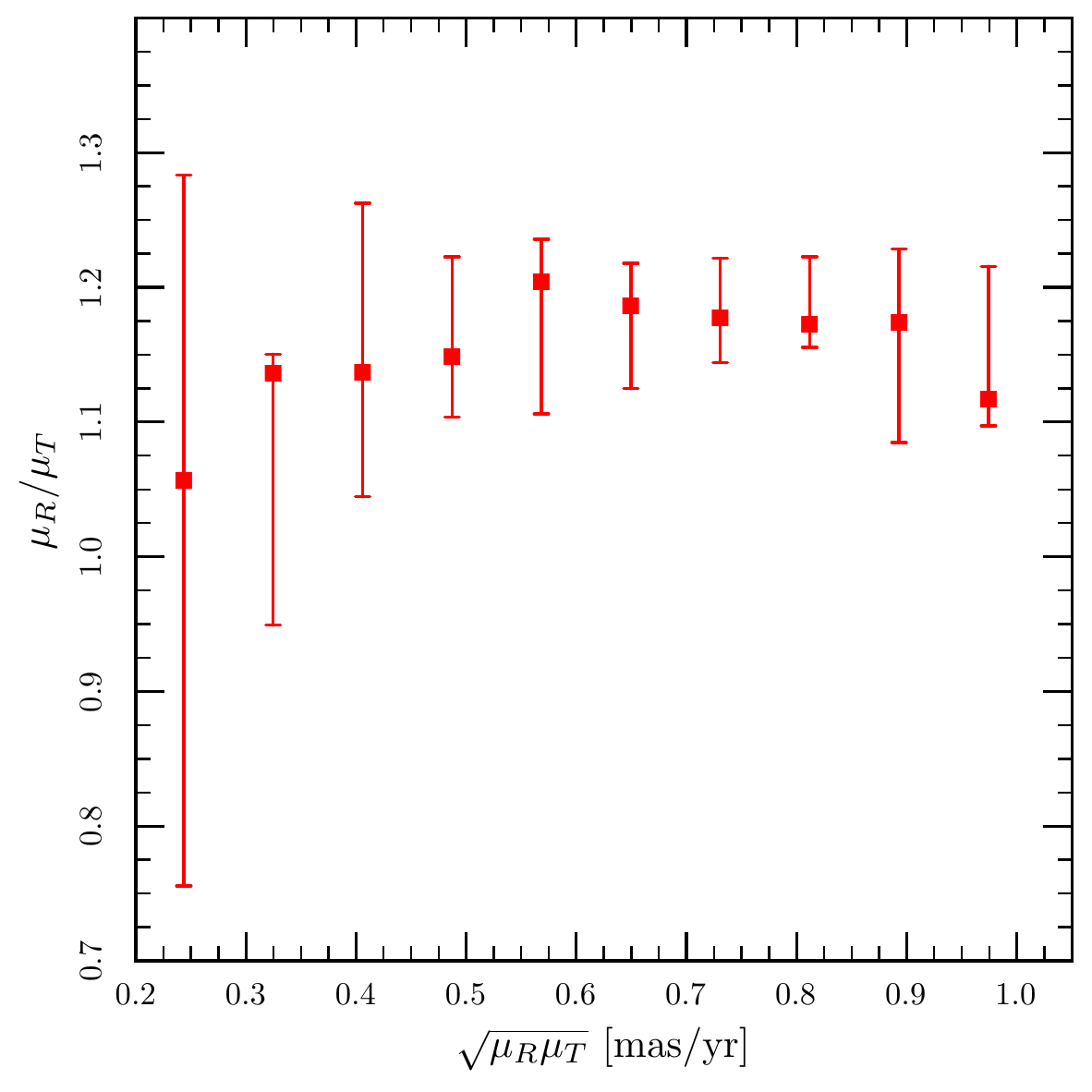}
\caption{The ratio of the major to the minor axis of the isodensity
  contours of phase-space density.}
\label{fig:contourratio}
\end{figure}

We would like to examine not only the proper motion distribution along
the radial and tangential directions but also the full two-dimensional
distribution.  To accomplish this we develop two new techniques to use
the proper-motion catalogues directly to obtain contours of constant
density (or isopleths) in proper motion.  These techniques are
outlined in the appendix.  The resulting contours for the
main-sequence stars in 47~Tuc with proper-motion errors less than
twice the median proper-motion error for stars of their magnitude are
depicted in Figure~\ref{fig:drawcontourr}.  For small values of the
proper motion, the gradient in the phase-space density of stars is
small, so the shape and centres of the ellipses are poorly
constrained.  In general the shape of the proper-motion ellipses does
not depend on their area (Figure~\ref{fig:contourratio}).  The
distribution of high-velocity stars has a similar anisotropy to that
of the low-velocity stars.  Furthermore, the contours of phase-space
density are well aligned with the radial and tangential directions as
expected from the theoretical models
(\S\ref{sec:modelling-47-tucanae}).  However, the location of the
centers of the contours does appear to depend on the area of the
contour.  The larger contours are centered on more positive proper
motions in the tangential direction than the smaller contours.  This
results from the skewness of the velocity distribution in the
tangential direction.  This skewness is a signature of the rotation.
It necessarily vanishes in the non-rotating models.

\begin{figure}
\includegraphics[width=\columnwidth]{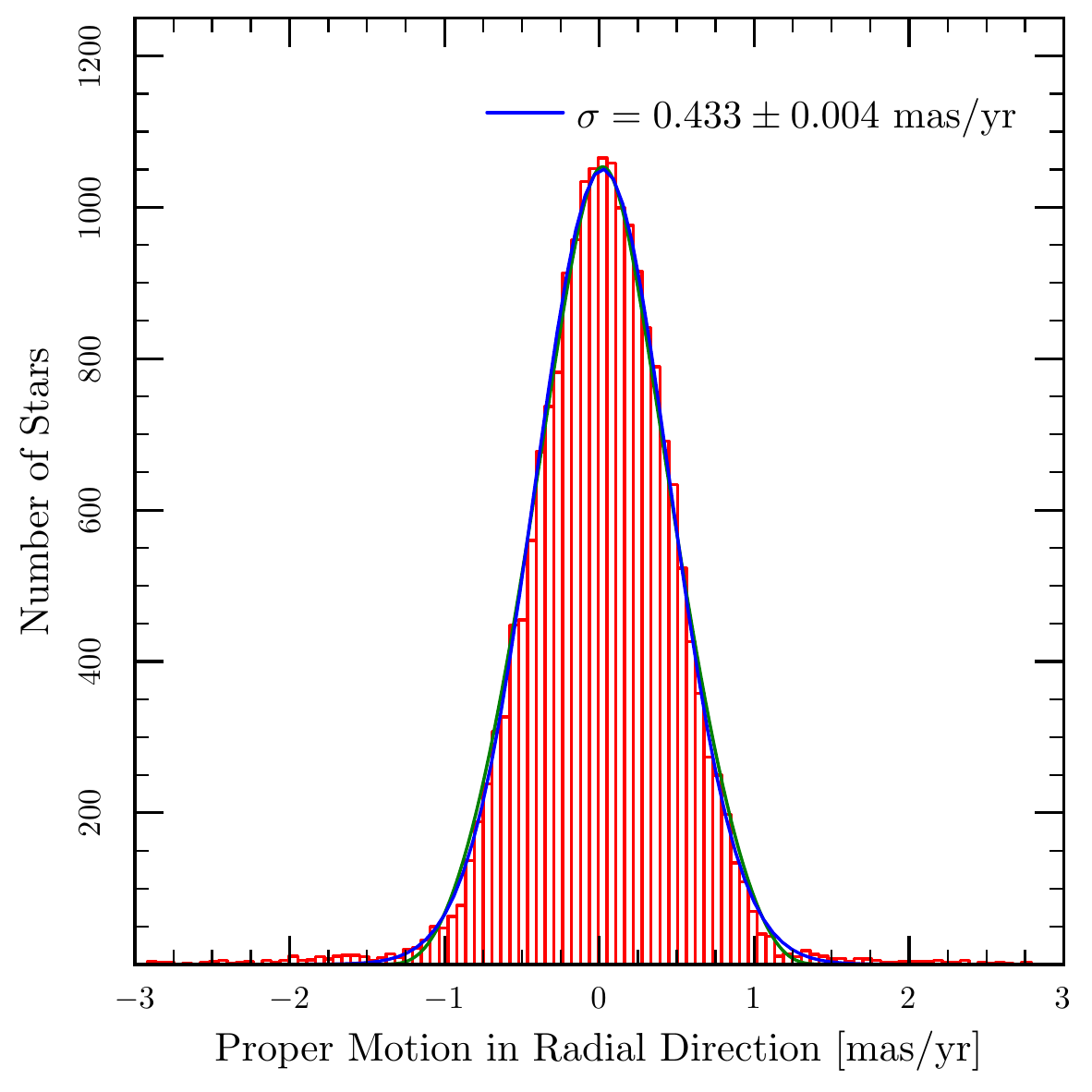}
\includegraphics[width=\columnwidth]{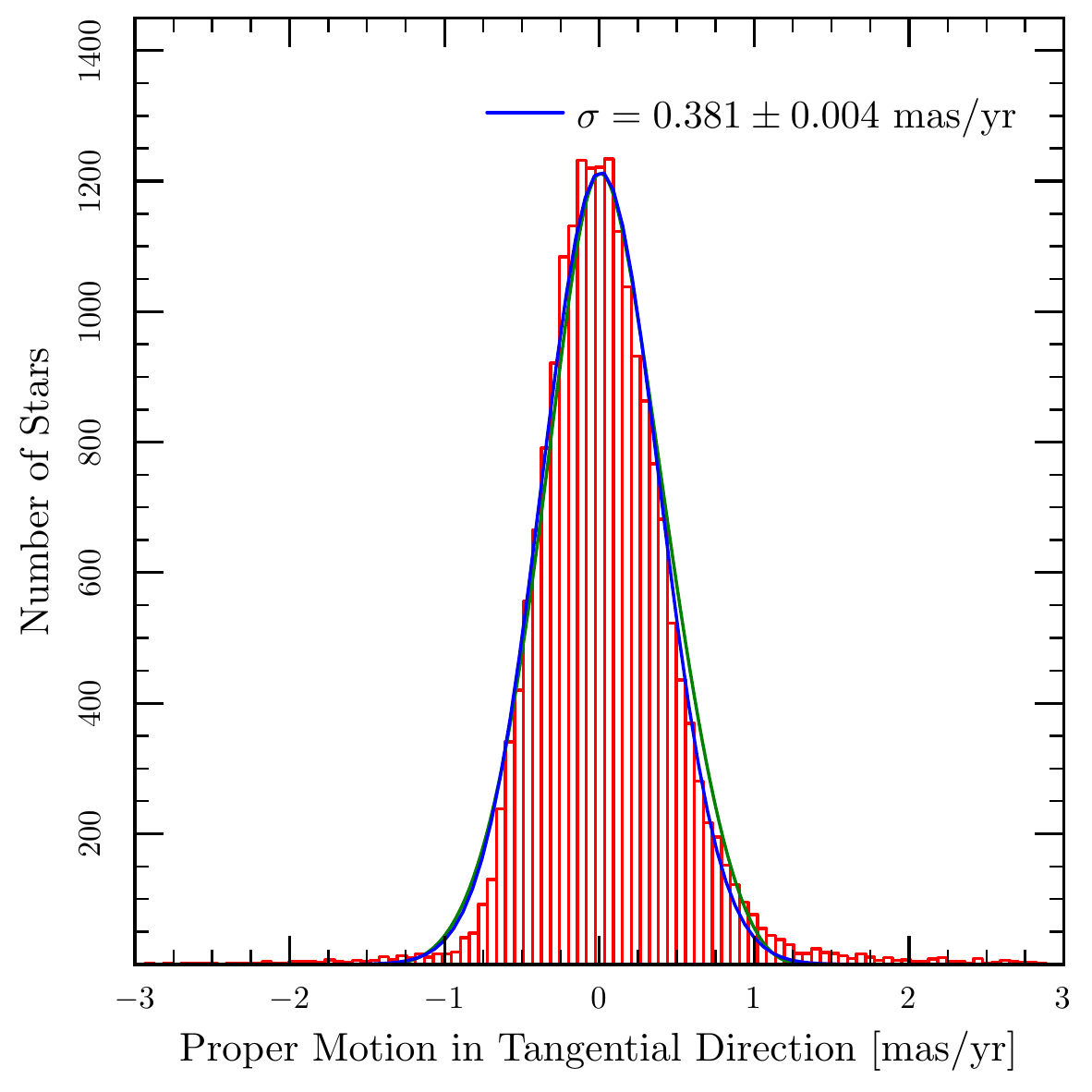}
\caption{Differential distributions of the observed proper motions in
  the radial and tangential directions with the best-fitting normal distributions (blue) and the distributions from the models (green).}
\label{fig:pm_dist}
\end{figure}

The ratio of major to minor axis of the elliptical isopleths is
approximately constant within the ninety-percent confidence regions
with the area of the ellipse, \ie with the typical size of the
proper motion.  We continue to examine the distributions in further
detail by looking at the one-dimensional distributions of the proper
motion along the radial and tangential directions.  The differential
distributions of the radial and tangential proper motions are shown in
Figure~\ref{fig:pm_dist}.  Superimposed on each histogram is the
best-fitting normal distribution with zero mean.  The normal
distribution is fit by minimizing the Kolmogorov-Smirnov statistic
between the normal distribution and the observed cumulative
distribution of proper motions.  The distribution of radial proper
motions as shown in the upper panel is well characterized by a normal
distribution with a dispersion of $\sigma = \left ( 0.433 \pm 0.004
\right ) \textrm{mas yr}^{-1}$.  We now focus on the tangential
distribution that exhibits a smaller dispersion of $\sigma = \left (
0.381 \pm 0.004 \right ) \textrm{mas yr}^{-1}$.  Comparing the normal
distribution with the observed differential distribution of tangential
proper motions shown in the lower panel reveals a few key differences.
Firstly, the observed distribution exhibits heavier tails than the
normal distribution (\ie it has positive kurtosis) and the model distributions (green) exhibit lighter tails than the normal distribution.  Secondly and more importantly, the observed distribution of tangential proper motions is skewed toward positive proper motions as we saw in Figure~\ref{fig:drawcontourr}. (Figure~\ref{fig:contourratio}).  The model (green curve) does exhibit a modest amount of skewness but not as large as the data.  

\begin{figure}
\includegraphics[width=\columnwidth]{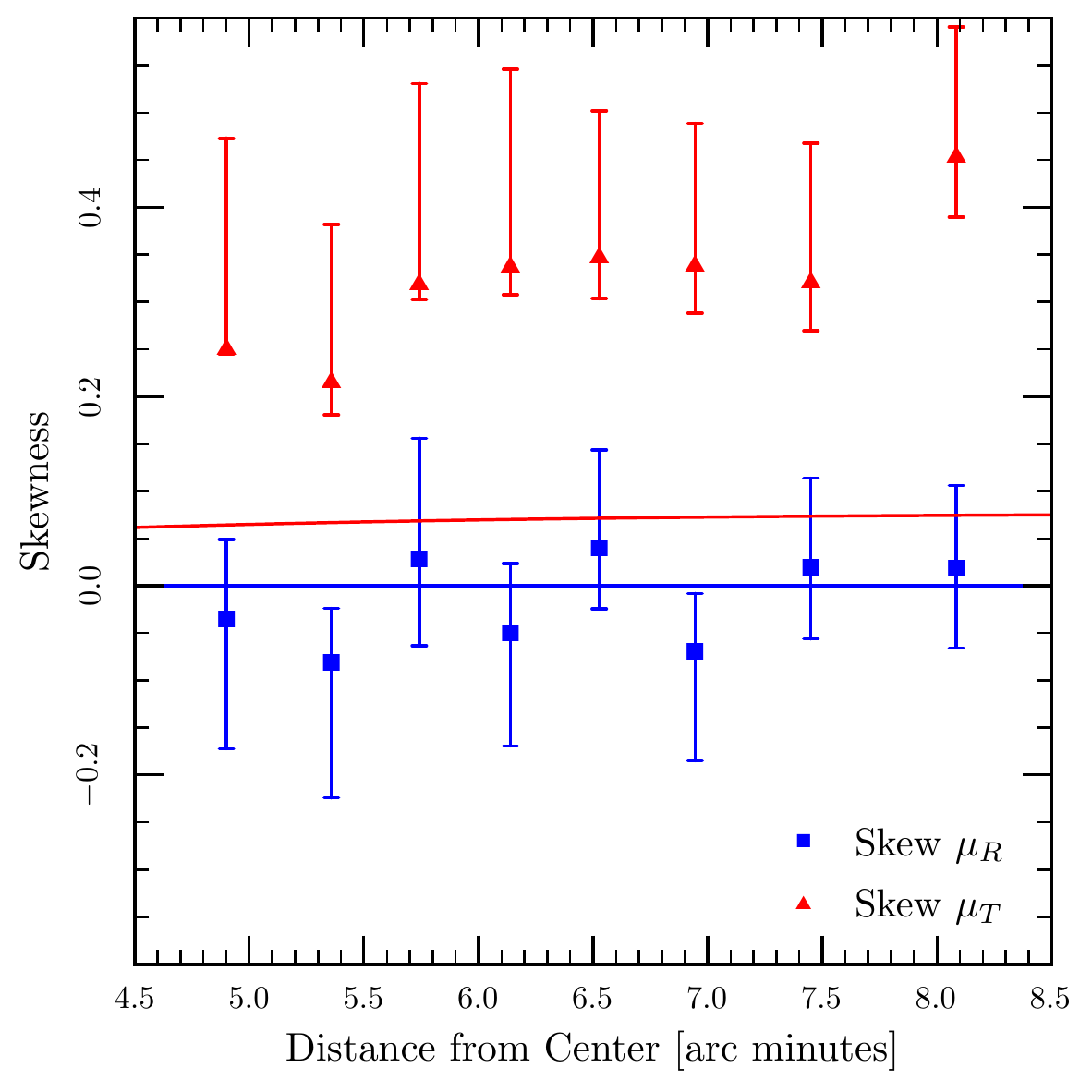}
\caption{
The value of the skewness as a function of projected radius.  The solid curves trace the results from the best-fitting rotating anisotropic model.  The observed skewness was {\bf not} used in the fitting process.}
\label{fig:skew_plot}
\end{figure}

We divided the stars into eight radial bins each with about 2,000 stars and performed an Anderson-Darling test to test the hypothesis that the the radial and tangential proper motions were drawn from a normal distribution.  For the radial distributions the $p-$values ranged from 0.01 to 0.81, indicating little evidence for a deviation from a normal distribution.  On the other hand, the $p-$values for tangential distributions ranged from $6 \times 10^{-17}$ to $3\times 10^{-4}$ indicating that these proper motions were unlikely to be drawn from a normal distribution.  To study this further we calculate the skewness of these distributions.  Figure~\ref{fig:skew_plot} examines the skewness of the radial and tangential proper motion distributions as a function of the projected radius from the center of the cluster.  We remove about 250 stars from the sample whose observed proper motions are likely to be larger than the local escape proper motion from the total sample of 16,881 stars before calculating the skewness. 
The skewness was not used in fitting the rotating, anisotropic model (\S~\ref{sec:model-47-tucanae}), the best-fitting model as shown by the solid lines exhibits a value of the skewness about a factor of four lower than that observed in the cluster proper motions.  To explore whether the large skewness exhibited by the data resulted from outliers, we also used the robust medcouple estimator of the skewness \citep{doi:10.1198/106186004X12632}, and we also found that observed skewness was typically larger than that of the models also by a factor of about three.
% Although the skewness was not used in fitting the rotating, anisotropic model (\S~\ref{sec:model-47-tucanae}), the best-fitting model as shown by the solid lines exhibits a value of the skewness consistent within the errorbars to that observed in the cluster.  Even changing the effects of rotation by 20\% yields models that are not consistent with the observed skewness.

%% How much is central velocity dispersion depressed?
%%
%% # Graph from /Users/heyl/Documents/Projects/FermiFAST/text/core_pmrradius_watk.png, page unknown
%%
%%                       ratio       ratio^2 - mass ratio
%%
%% 0.05389	0.63660 1.0766654828 1.1592085618
%% 0.05571	0.59127
%% 1.255	0.58810 1.0714155584 1.1479312988
%% 1.275	0.54890
%%

% Finally we revisit the two-dimensional distribution of proper motions. In particular because the shape of the ellipses are constant with the area of the ellipses, we examine the phase-space density as a function of the size of the ellipse in Figure~\ref{fig:contourdensity}.  The two normal distributions depicted in Figure~\ref{fig:pm_dist} are superimposed on the observed phase-space density.  The observed phase-space density is well characterized by 
% \begin{equation}
% f(\mu_R,\mu_T) d \mu_R d \mu_T \propto \exp \left ( -\frac{\mu_R
%     \mu_T}{2 \sigma^2} \right )  d \mu_R d \mu_T 
% \label{eq:5}
% \end{equation}
% where $\sigma = 0.395~\textrm{mas yr}^{-1}.$

% \begin{figure}
% \includegraphics[width=\columnwidth]{contourrdensity}
% \caption{The phase-space density of stars along the ellipsoidal
%   contours as a function of the square-root of the area of the
%   contour.
% }
% \label{fig:contourdensity}
% \end{figure}

\subsection{The Mass of 47~Tucanae}
\label{sec:mass-47-tucanae}

\citet{1989ApJ...3339..195L} outlined several mass estimators for the open cluster M35 using the observed proper-motions.  Among these the most straightforward is
\begin{equation}
\label{eq:6}
\langle G M_r \rangle = \frac{16}{\pi} \left \langle R \left (
    \frac{2}{3} v_R^2 + \frac{1}{3} v_T^2 \right ) \right \rangle
\label{eq:7}
\end{equation}
where $v_R$ is the component of the velocity toward the center of the cluster and $v_T$ is the tangential velocity. Both of these are measured in the plane of the sky. This estimator yields
\begin{equation} 
  \label{eq:8}
%  M_r = 4.1 \pm 0.2 \times 10^4 d_2^3 \mathrm{M}_\odot
  M_r = 8.36_{-0.16}^{+0.18} \times 10^5 d_{4.7}^3 \mathrm{M}_\odot
\label{eq:9}
\end{equation}
within 6.4~arcminutes or $1.28_{0.02}^{+0.03} \times 10^6 d_{4.7}^3 \mathrm{M}_\odot$ for the entire cluster using the best-fitting anisotropic model to scale the result of Eq.~\ref{eq:8} out to the tidal radius.  We have summed over all of the stars along the cluster main sequence and excluded those stars with proper motion errors greater than 0.4~mas/yr or proper-motions greater than 5~mas/yr.  The confidence interval gives a ninety-percent confidence region obtained through bootstrapping the sample.

By comparing the motion of the stars in two fields on either side of the center of 47~Tuc with the motion of the stars in the Small Magellanic Cloud, \citet{2003AJ....126..772A} estimated the rotational velocity in this field to be about $v_\mathrm{rot} = 0.267~\mathrm{mas~yr}^{-1}$, yielding an addition to the mass estimator,
\begin{equation}
\langle G M_r \rangle = \frac{16}{\pi} \left \langle R \left [
    \frac{2}{3} v_R^2 + \frac{1}{3} \left ( v_T^2 + v_\mathrm{rot}^2
    \right ) \right ] \right \rangle,
\label{eq:10}
\end{equation}
and a slightly larger mass of 
\begin{equation}
  \label{eq:11}
%  M_r = 4.1 \pm 0.2 \times 10^4 d_2^3 \mathrm{M}_\odot
  M_r = \left ( 9.61 \pm 0.16 \right ) \times 10^5 d_{4.7}^3 \mathrm{M}_\odot
  \label{eq:11}
\end{equation}
% (0.5794267297 model units - phi = 1.867 - v_e )
within 6.4~arcminutes or $\left ( 1.31 \pm 0.02 \right ) \times 10^6
d_{4.7}^3 \mathrm{M}_\odot$ for the entire cluster using the
best-fitting rotating anisotropic model to scale to the tidal radius.
Using the fit of the best-fitting rotating anisotropic model to the
proper motions yields a larger mass estimate of $1.4\times
10^6~\mathrm{M}_\odot$ (from Table~\ref{tab:model_data}). 
% In this
% global model the mass column density throughout the cluster is
% estimated using the photometry as depicted in the upper panel of
% Figure~\ref{fig:best_models} using a constant mass-to-light ratio.  Mass
% segregation toward the core will decrease the mass-to-light ratio of
% the core relative to the outskirts and possibly produce this 22\%
% overestimate of the total mass \citep[\eg][]{2013arXiv1308.3706G}
% relative to the dynamical mass estimate from Eq.~\ref{eq:10}.

% We can also estimate the mass within the distance of our field from the center of the % cluster using Jeans equation
% \begin{equation}
% v_\mathrm{rot}^2 - \sigma_r^2 \left [ \frac{d \ln \sigma_r^2}{d \ln r} + \frac{d
%    \ln n}{d \ln r} + \beta \right ] = \frac{G M_r}{r}
% \end{equation}
% where
% \begin{equation}
% \beta(r) = 1 - \frac{\sigma^2_\phi(r)}{\sigma^2_r(r)}
% \end{equation}

We can compare the more conservative total mass estimate of $\left (
1.31 \pm 0.02 \right ) \times 10^6 d_{4.7}^3 \mathrm{M}_\odot$ with
the observed luminosity of the cluster of $5 \times 10^5 d_{4.7}^2
\mathrm{L}_{V,\odot}$ to obtain a mass-to-light ratio of $2.6 d_{4.7}
\mathrm{M}_{\odot}/\mathrm{L}_{V,\odot}$. This is similar to the value
of $2.4 d_{2.53} \mathrm{M}_{\odot}/\mathrm{L}_{V,\odot}$ that we
found from a similar analysis in NGC~6397 \citep{Heyl116397dyn}.

The stellar density and velocity dispersion can provide an estimate of the relaxation time for a group of stars \citep{1971ApJ...164..399S},
\begin{equation}
\tau_r = \frac{0.065 v_\mathrm{RMS}^3}{3 G^2 m_* \rho \ln (0.4 M/m_*)}
\end{equation}
where $v_\mathrm{RMS}$ is root-mean-square velocity of the stars, $\rho$ is the mass density, $M$ is the total mass of the cluster and $m_*$ is the mass of a typical star (taken to be 0.39~M$_\odot$).  Using Eq.~\ref{eq:11} and the information in Table~\ref{tab:rmserr}, we obtain
\begin{equation}
\tau_r = 2.6~d_{4.7}^6 \times 10^{10}~\mathrm{years}.
\end{equation}

\subsection{Stellar Escapers}
\label{sec:stellar-escapers}

\begin{figure}
\includegraphics[width=\columnwidth]{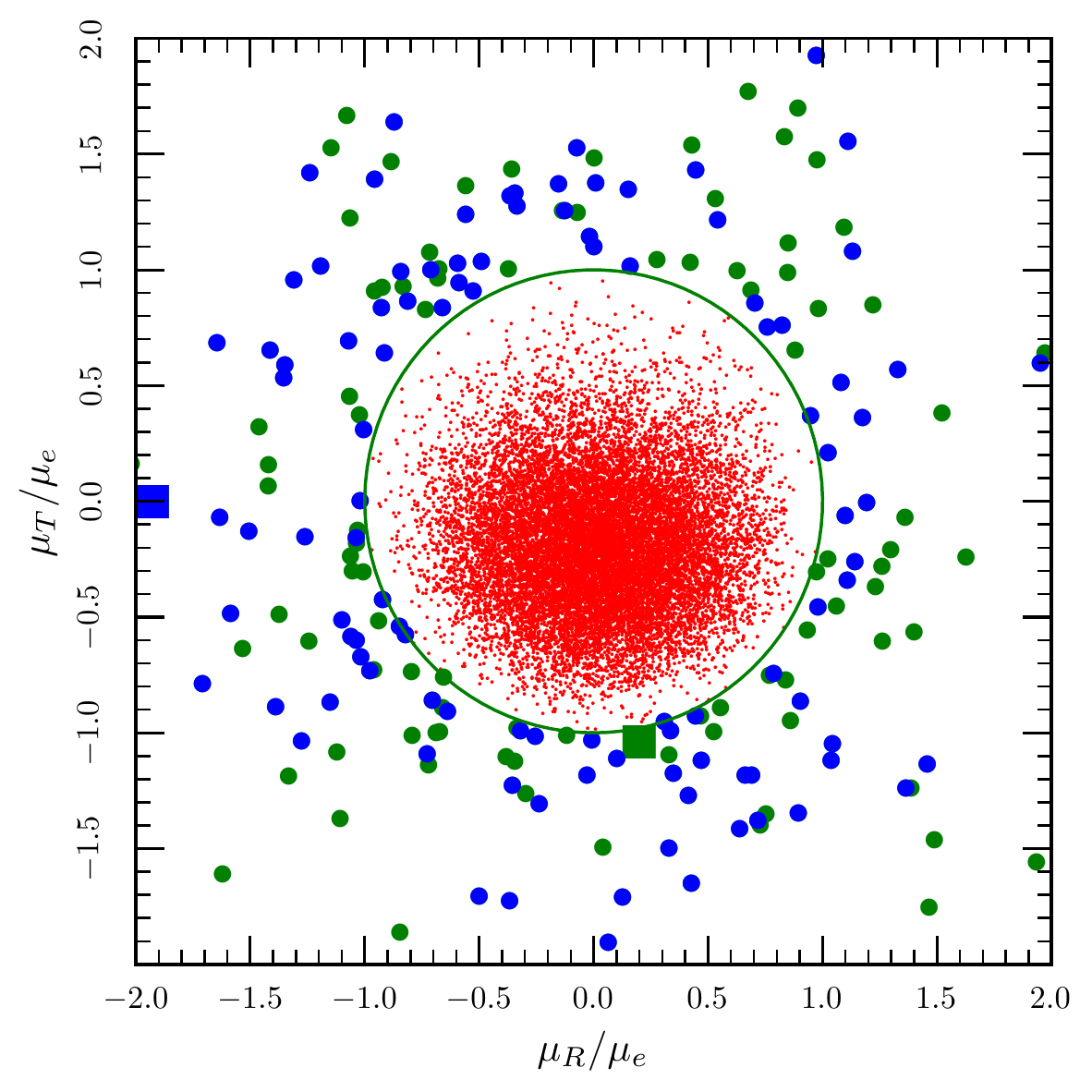}
\includegraphics[width=\columnwidth]{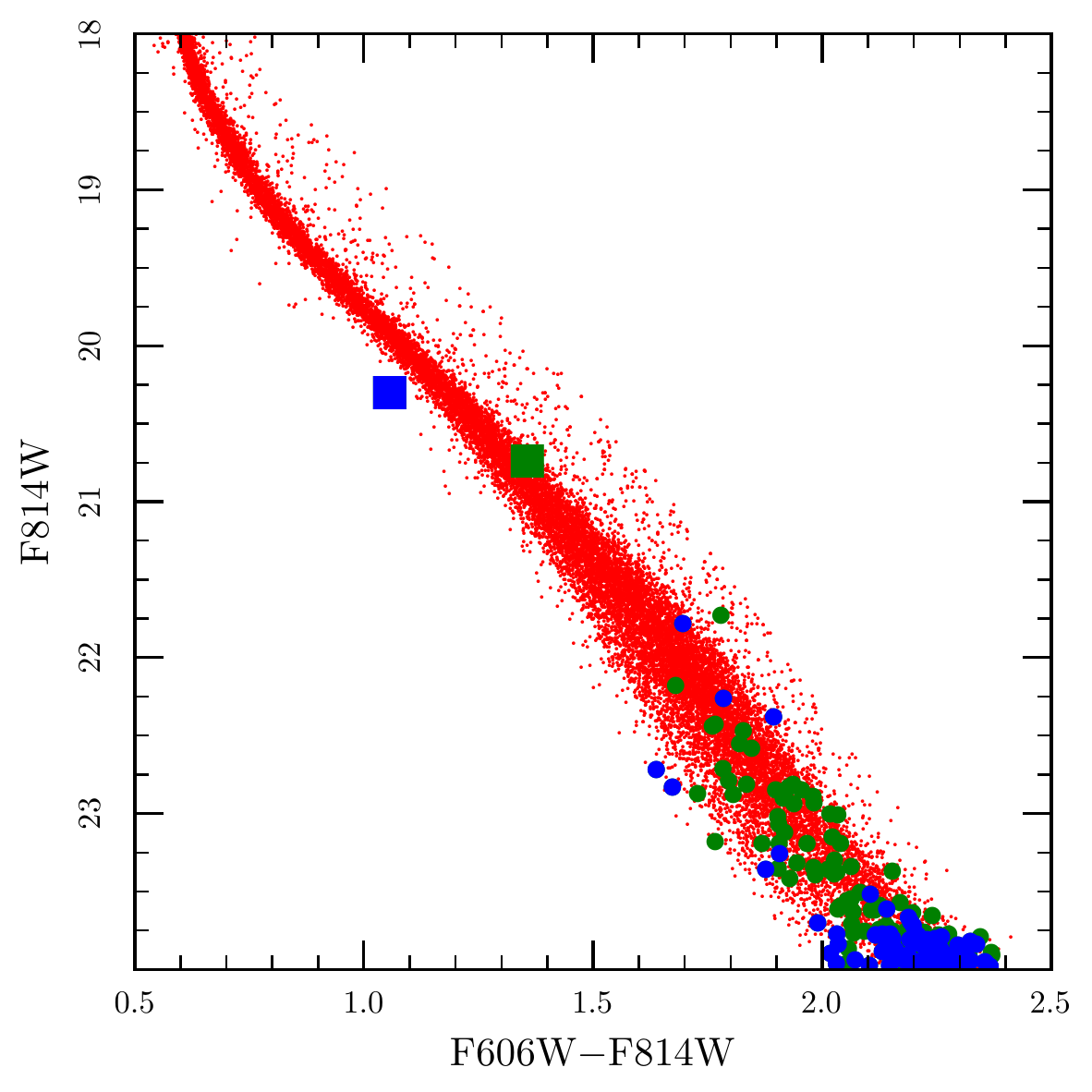}
\caption{Upper Panel: The Proper Motions of Stars along the Main
  Sequence as Compared to the Escape Proper Motion.  Lower Panel: The
  Location on the Colour-Magnitude Diagram of the Potential Escapers.
  The blue symbols depict stars with proper-motion errors less than
  the median proper-motion error at the magnitude of the star.  The
  green symbols depict stars with proper-motion errors greater than
  the median proper-motion error.  The larger squares are stars whose
  F814W magnitude lies between 18 and 21 where the proper motions are
  best determined.}
\label{fig:pmescape}
\end{figure}

Using the rotating, anisotropic model of the cluster, we determine the
best power-law fit to the escape proper motion as a function of projected
radius as
\begin{equation}
v_e = 1.4 \left ( \frac{R}{5~\textrm{arcminutes}} \right )^{-0.4}
\textrm{mas yr}^{-1}
\label{eq:18}
\end{equation}
where we have assumed that each star has no radial velocity component and that each lies at the tangent point, \ie $r=R$, to yield a conservative estimate of the escape
velocity for each star.  The upper panel of Figure~\ref{fig:pmescape}
depicts the proper motions of the stars compared to the escape proper
motion, and the lower panel gives their locations on the
color-magnitude diagram.  We have added the rotational velocity in
the plane of sky as observed by \citet{2003AJ....126..772A} to
calculate the total proper motion of the stars. At the faintest
magnitudes there are many potential escapers.  However, it is at these
magnitudes that the proper-motion error estimates are suspect (see
Figure~\ref{fig:pmrmag}), so we shall focus on the 8,700 stars with $18
< \textrm{F814W} < 21$.  In this magnitude range there are only two
potential escapers. One of them is just barely an escaper and its
proper motion error is larger than the median.  The other star clearly
has a velocity much larger than the escape velocity.  However, on the
color magnitude diagram it lies at the intersection of the giant
branch of the SMC with the main sequence of 47~Tuc, so it may be a member of the SMC rather than 47 Tuc.  
% If we use the
% presence of a single escaper with the crossing time at this radius
% (0.7~Myr), we obtain an estimate of the evaporation time of about six
% billion years for the stars in this field.  Because the actual
% association of the star with 47~Tuc is uncertain, this is a lower
% limit on the evaporation time. 

\subsection{Central Proper Motions}
\label{sec:centr-prop-moti}

% Central (83.3333)
% mean 47 Tuc: 2.95003e-05 -3.66757e-06
% mean SMC     0.0582708 0.0148969
%
% Outer (50)
% mean 47 Tuc 0.000583324 0.000736651
% mean SMC 0.0944338 0.0280963

\begin{figure}
  \includegraphics[width=\columnwidth]{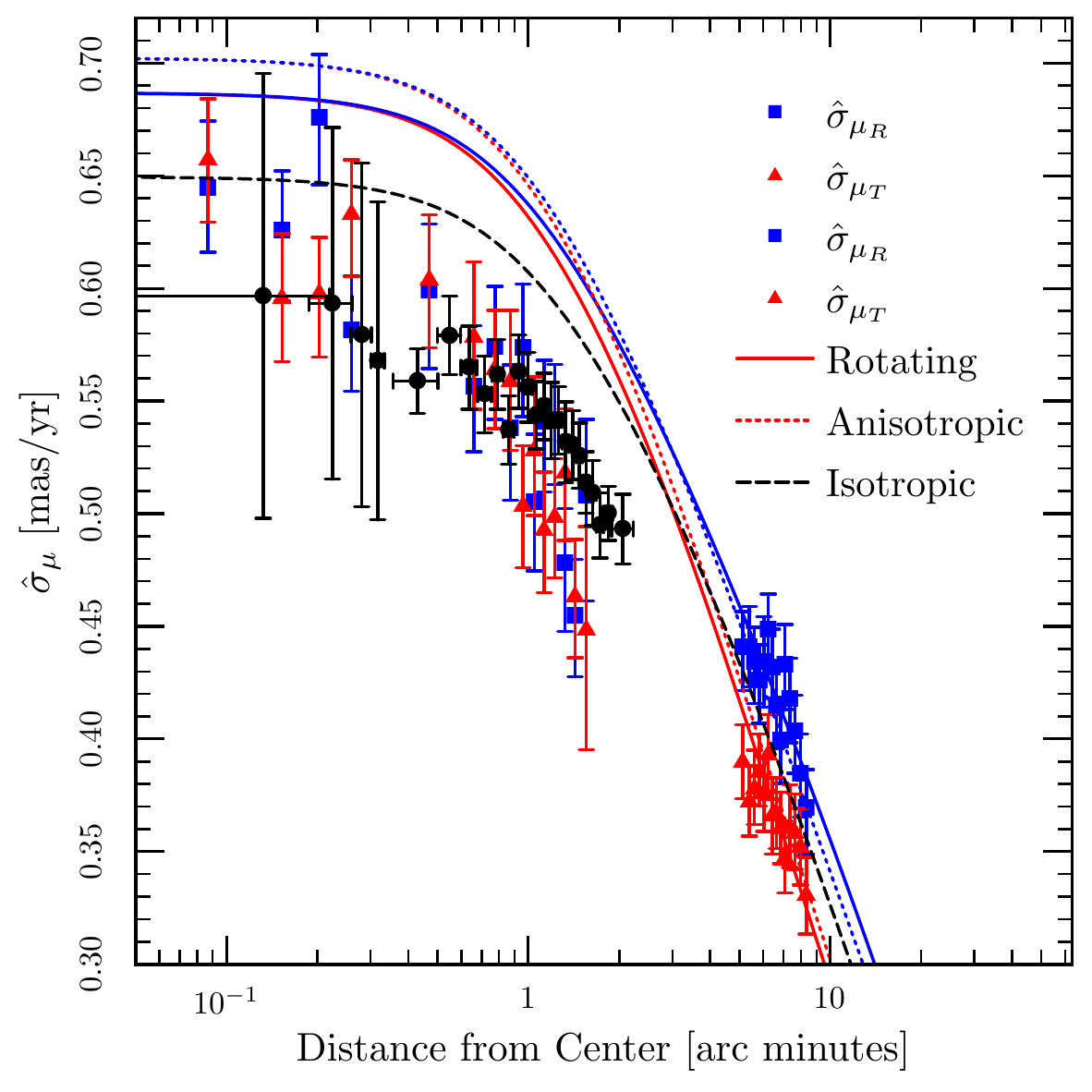}
   \includegraphics[width=\columnwidth]{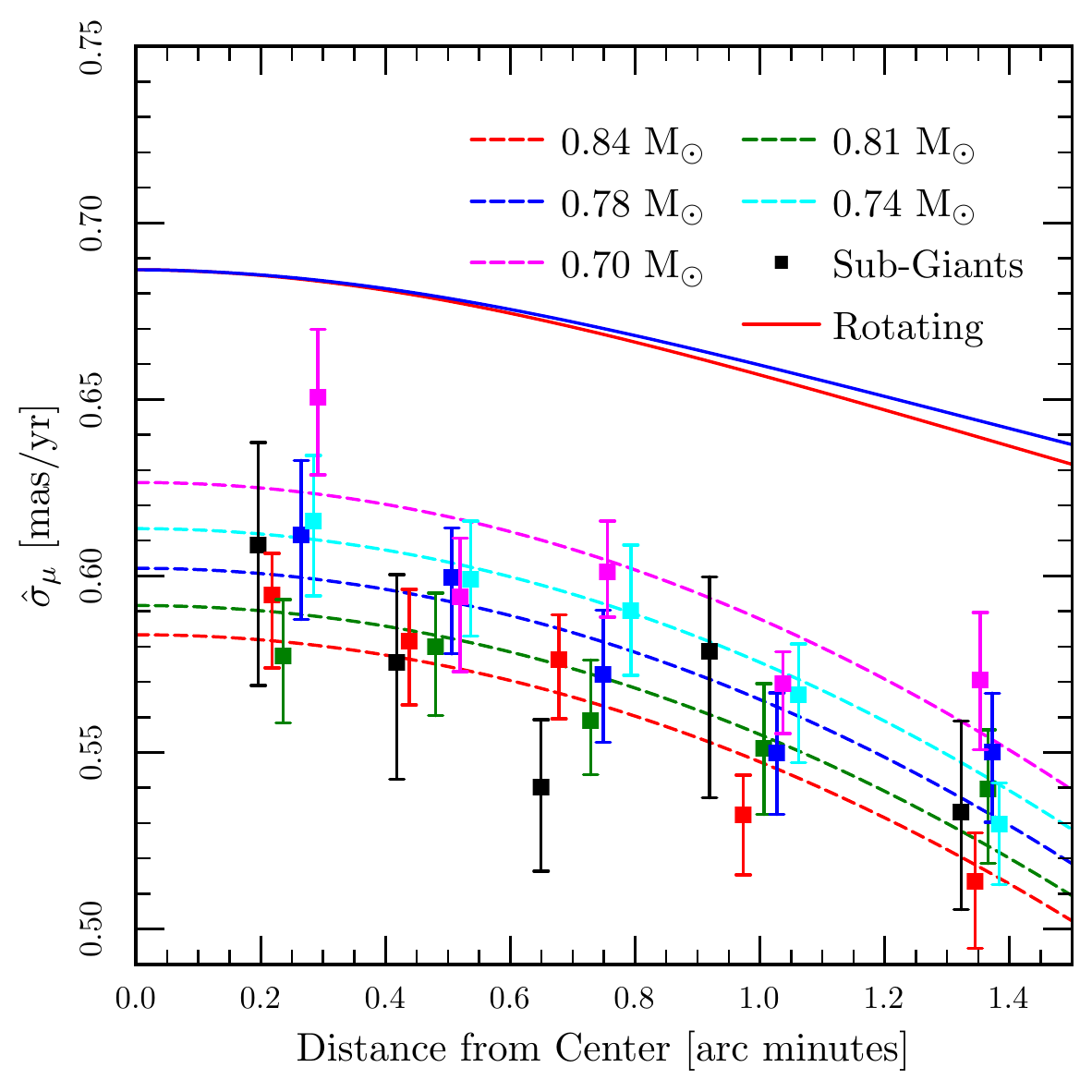}
\caption{Upper panel: The velocity
  dispersions outside of three arcminutes are from this paper while
  those within three arcminutes are from \citet{2006ApJS..166..249M}
  in blue and red and \citet{2015ApJ...803...29W} in black.  Both
  are depicted with ninety-percent confidence regions determined by
  bootstrapping for our data and by scaling the one-sigma errorbar
  s
  for the \citet{2006ApJS..166..249M} and \citet{2015ApJ...803...29W} data.
  Lower panel: Inner proper motions from our WFC3 UV data with the
  global models and mass-segregation fits for comparison.
}
\label{fig:coredispersions}
\end{figure}

Because of the tendency toward kinetic energy equipartition, more massive stars have smaller velocity dispersion than low mass ones and a more concentrated distribution \citep{2013arXiv1308.3706G,Heyl14diff,2016arXiv160505740P}. This effect, not accounted in our single-mass models, is expected to be particularly important in the central regions of the cluster. Figure~\ref{fig:coredispersions} shows the velocity dispersions measured by \citet{2006ApJS..166..249M} and \citet{2015ApJ...803...29W}.  Our rotating model and our isotropic model overpredict the central velocity dispersion.  The overprediction by the non-rotating anisotropic model is more dramatic. The range of masses probed by the \citet{2006ApJS..166..249M} data is 0.65 to 0.9~solar masses and by the \citet{2015ApJ...803...29W} data is 0.8 to 0.9~solar masses, significantly larger than the median mass of our outer dataset of about 0.4 solar masses. 
%We have plotted an illustrative rotating model compare by scaling the proper motion dispersion to account for the ratio of the median masses of the two samples using the mass scaling in the outer field discussed in \S~\ref{sec:prop-moti-disp}, Eq.~\ref{eq:3}~and~\ref{eq:4}. Of course, this model is not internally consistent, but it allows use to compare the data in the two regions. 
The observed velocity dispersions in the core are smaller than those expected from the model fitted with the outer model by a factor similar to that given by the outer-mass-segregation formulae discussed in \S~\ref{sec:prop-moti-disp}, Eq.~\ref{eq:3} and~\ref{eq:4}.  The anisotropic and rotating anisotropic models both predict the anisotropy in the observed proper motions and of course the isotropic model does not.  However, the anisotropy both in the models and the data is mild in the central region.  The observed central velocity dispersions were not used in the fits, but are just presented for comparison.

Because our central field is centered on the center of the cluster, we can measure the mean proper motion of the central stars and compare it to the mean proper motion of the stars in the outer field, using the stars of the Small Magellanic Cloud as a reference in both cases to obtain an estimate of the rotation of the clusters in the outer field relative to the inner field \citep{2003AJ....126..772A,2017ApJ...844..167B}.  We find that the median tangential proper motion of the stars in the outer field is $0.28 \pm 0.04~\mathrm{mas~yr}^{-1}$ with ninety percent confidence in agreement with the previous results.  On the other hand, the best fitting rotating model yields a median tangential proper motion in the field of $0.08~\mathrm{mas~yr}^{-1}$.  Similar to the local measurements of skewness the best-fitting Lupton-Gunn model falls short in the determining the rotation of the cluster.  Even models that had values of $\omega$ four times larger than the model that fit the surface brightness and velocity dispersions best still could not yield values of the skewness or rotation rate as large as observed; furthermore, these models did not fit the surface brightness distribution and velocity dispersions nearly as well as the more slowly rotating models.

We examine the central velocity dispersions further using our WFC3 data in the core of 47~Tucanae.  We divide the stars with proper motions along the main sequence into five magnitude bins each with equal numbers of stars.  We only use stars with $\mathrm{F336W}>18.1$ so we exclude the subgiant branch from our main-sequence sample. Additionally we examine an independent sample of sub-giant stars.  The magnitude limits of the bins are summarized in the lower portion of Table~\ref{tab:model_data}.  Furthermore, we divide each of these magnitude bins into five radial bins again with equal numbers of stars.  We perform an Anderson-Darling test on the radial and tangential proper motions within each of these 25 bins and find that the $p-$values for 49 of the 50 samples exceed 0.01, indicating that there is little evidence that the distributions deviate from normality; therefore, we calculate the proper motion dispersion in each bin for those stars whose proper motion errors are smaller than the median proper motion error.  We have combined the radial and tangential proper motions in the dispersion calculation as they do not differ from each other significantly. The results of this analysis are depicted in the lower panel of Figure~\ref{fig:coredispersions} along with the rotating anisotropic models.

We see that the proper motion dispersion depends not only on radius but also on the magnitude of the bin.  We use the MESA models discussed in \S~\ref{sec:observations} to estimate the masses of the stars in each bin and fit the observed proper motion dispersions with a formula of the form
\begin{equation}
  \hat \sigma_\mu = \left [a - b \left (
    \frac{r}{1~\mathrm{arcminute}} \right )^2 \right ] \left (
  \frac{M}{\mathrm{M}_\odot} \right )^c.
  \label{eq:19}
\end{equation}
This is not a self-consistent model just an empirical fitting relation.
We find $a=0.54$~mas/yr, $b=0.034$ and $c=-0.4$, and the resulting
fitted curves are depicted as dashed lines in the figure.  This is
somewhat weaker than expected from the observed mass segregation in
radius \citep{2013arXiv1308.3706G} which would yield $c\approx-0.48$.
It is also somewhat weaker than energy equipartition with $c=-0.5$.
From Table~\ref{tab:rmserr} we see that the median mass of the least
massive bin in the core is $0.7~\mathrm{M}_\odot$ whereas the median
mass of the stars used to constrain the global model is
$0.38~\mathrm{M}_\odot.$

We can place these measurements of the mass segregation in the two fields in the context of the results of \citet{2016MNRAS.458.3644B} who found that the velocity dispersion measured in numerical models of globular clusters is well characterized by a function of the form
\begin{equation}
\sigma = \sigma_\mathrm{eq} \times \left \{ 
\begin{array}{cl}
\exp \left [ \frac{1}{2} \frac{m_\mathrm{eq}-m}{m_\mathrm{eq}} \right ] & \mathrm{if}~m < m_\mathrm{eq} \\
\left ( \frac{m}{m_\mathrm{eq}} \right )^{-1/2}
 & \mathrm{if}~m > m_\mathrm{eq}
\end{array} 
\right . .
\label{eq:bianchini_1}
\end{equation}
More crucially for our purposes the relation exhibits the following logarithmic derivative of velocity dispersion with mass
\begin{equation}
\frac{d \ln \sigma}{d \ln M} = \left \{ 
\begin{array}{cl}
-\frac{1}{2} \frac{m}{m_\mathrm{eq}} & \mathrm{if}~m < m_\mathrm{eq} \\
-\frac{1}{2} & \mathrm{if}~m > m_\mathrm{eq} 
\end{array} 
\right . .
\label{eq:bianchini_2}
\end{equation}
In both fields the power-law dependence with mass is weaker than $-0.5$.  This yields an estimate of the value of $m_\mathrm{eq}$ in the two fields.  In the outer field we have $d \ln \sigma / d \ln M \approx -0.1$ and a median mass of 0.4~solar masses, giving $m_\mathrm{eq} \approx 2~\mathrm{M}_\odot$.  In the inner field, the typical masses are about two times larger but the power-law index is four times larger, yielding $m_\mathrm{eq}$ of about one solar mass.

Furthermore, we can compare the proper motion dispersion near the
center of the cluster with the radial proper motion dispersion
measured by \citet{2006ApJS..166..249M} of $11.6 \pm
1.3~\mathrm{km~s}^{-1}$ (ninety-percent confidence).
\citet{2006ApJS..166..249M} restricted their radial velocity sample to
stars with $11 < V < 14$ and projected distances from the center of
less than 105 arcseconds.  This sample of stars is dominated by the
horizontal branch stars whose mass is about
$0.905~\mathrm{M}_\odot$ according to our stellar models.  We assume
following the results of \citet{Heyl14massloss} and
\citet{2016arXiv160505740P} that little mass is lost before the
horizontal branch.  We use the entire sample of subgiants to determine
the one-dimensional proper motion dispersion of $0.57\pm0.01$~mas/yr
with a ninety-percent confidence interval in agreement with the results of \citet{2015ApJ...803...29W}.  The mass of this sample is
about $0.89~\mathrm{M}_\odot$, only slightly lower than that of the horizontal
branch stars, so the expected proper motion dispersion of the
horizontal branch stars would only be smaller by one percent if we use
the measured scaling of Eq.~\ref{eq:19}.  This yields a kinematic
distance of $4.32\pm0.47$~kpc with ninety-percent confidence where the
uncertainty is dominated by the published uncertainties in the
measurement of the radial velocity dispersion.  On the other hand if we assume that the horizontal branch stars have a mass of $0.7~\mathrm{M}_\odot$ \citep{2016A&A...590A..64S}, we can use the stars of Bin 5 in Table~\ref{tab:rmserr} to estimate the proper motion dispersion of this population at $0.59 \pm 0.01$~mas/yr yielding a distance of $4.15\pm0.45$~kpc.

\section{Discussion}
\label{sec:discussion}

We have used a catalog of over 60,000 proper motions in the globular cluster 47~Tucanae to develop a global dynamical model of the cluster and examine the detailed properties of the observed distributions as a function of position and stellar mass.  Our sample consists both of proper motions in the core and proper motions in a field well beyond the half-light radius.  Our work in NGC~6397 \citep{Heyl116397dyn} and \citet{2012A&A...540A..94V} in omega Centauri are other studies of a globular cluster that have analyzed proper motions in an outlying field, and this study complements the recent study of \citet{2017ApJ...844..167B} that also combines observations in the core and outlying fields for 47~Tuc.  In NGC~6397 we did not find strong evidence for anisotropy even though this is expected to more evident further out in the cluster.  This agreed with our measurement of the mean relaxation time of stars the field of about 1~Gyr. The much richer globular cluster 47~Tuc results in stars with much longer relaxation times in our outer field and both anisotropy and rotation are manifest in the proper-motion distribution.

\subsection{Dynamics}
\label{sec:dynamics}

\citet{2013ApJ...772...67B} fit the rotating models of \citet{2012A&A...540A..94V} to 47~Tuc but their proper motion sample only extended to 1.5~arcminutes from the center, so they do not see evidence of anisotropy or rotation within the proper motions.  Subsequently \citet{2017ApJ...844..167B} fit rotating models to the proper motions in several fields with 47 Tuc.  In our samples the anisotropy and rotation in our outer ACS field are apparent from the proper motion diagram, Figure~\ref{fig:jay_pmall_paper}.  On the other hand within the central WFC3, both anisotropy and the effects of rotation are apparently absent in Figure~\ref{fig:cen_pmall}.  To account for both the rotation and the anisotropy we use a Lupton-Gunn (\citeyear{1987AJ.....93.1106L}) rotating globular cluster model to describe the cluster.  This model best accounts for both the proper-motion dispersion in our outer ACS field and for the surface brightness profile.   We did not use the skewness of the proper-motion distribution to fit the models; we find that the best-fitting model generates a skewed distribution of tangential velocities but the skewness and the mean of the observed velocities are larger than that in the models.  Increasing the assumed inclination of the model would reduce the predicted mean rotational velocity in the plane of the sky not increase it, so we conclude the cluster is rotating too fast for the Lupton-Gunn  model to be applicable at this level of precision.  We found that the proper-motion dispersion depends weakly on stellar mass with $\hat{\sigma}_\mu \propto M^{-0.10}$.  

When this model is extended into the center of the cluster, we find that the observed proper motions both in the literature and our measurements (Figure~\ref{fig:coredispersions}) are smaller than the expectations from the model.  The median mass of the stars in our inner WFC3 proper motion sample are about twice that in our outer ACS field.  We argue that the mild mass segregation observed in our outer field could be responsible for the observed decrease in the central proper-motion dispersion relative to the models.  Furthermore, we find that the core proper motions depend more strongly on mass with $\hat{\sigma}_\mu \propto M^{-0.40}$.  Here the relaxation time has been measured to be about 35~Myr \citep{Heyl14diff}, so we expect the proper motions to be more relaxed here than in the outer field.  

\subsection{Properties}
\label{sec:properties}

Because the ACS field lies well beyond the half-light radius of the cluster, the measurements yield an estimate for the total mass of the globular cluster with only modest extrapolation.  Our mass even as measured in the model independent dynamical estimators, $\left ( 9.61 \pm 0.16 \right ) \times 10^5 d_{4.7}^3 \mathrm{M}_\odot$ within 6.4~arcminutes, is larger than that in the literature \citep[\eg][ $\left (6.23 \pm 0.04 \right) \times 10^5 \mathrm{M}_\odot$]{2013ApJ...772...67B} for the entire cluster.  The total mass that we estimate for the cluster of  $\left ( 1.31 \pm 0.02 \right ) \times 10^6 d_{4.7}^3 \mathrm{M}_\odot$ yields a mass-to-light ratio of $2.6 d_{4.7} \mathrm{M}_{\odot}/\mathrm{L}_{V,\odot}$. This is similar to the value of $2.4 d_{2.53} \mathrm{M}_{\odot}/\mathrm{L}_{V,\odot}$ that we found from a similar analysis in NGC~6397 \citep{Heyl116397dyn}.  However, it is much larger than the value determined by \citet{2013ApJ...772...67B} of $1.69 \pm 0.13 \mathrm{M}_\odot/\mathrm{L}_{\odot,V}$.  If we scale our mass and mass-to-light ratio to the kinematic distance of \citet{2013ApJ...772...67B} of 4.1~kpc, we obtain $\left ( 8.6 \pm 0.02 \right ) \times 10^6 d_{4.1}^3 \mathrm{M}_\odot$ yields a mass-to-light ratio of $2.5 d_{4.1} \mathrm{M}_{\odot}/\mathrm{L}_{V,\odot}$, still substantially larger than their values.  Their result did not include kinematics at large distances from the center of 47~Tuc, but rather requires an extrapolation of the mass density observed in the core to large radii using a model constrained by the surface brightness or stellar column density. 
% \citet{2013ApJ...772...67B} also measured the mass-to-light ratio of omega Centauri.  In this case they use proper motion data both in the core and well outside the half-light radius, and they obtain a value of $2.86 \pm 0.14\mathrm{M}_\odot/\mathrm{L}_{\odot,V}$ close to our value in 47~Tuc.  This highlights the importance of kinematic data in outlying fields to constrain the mass of globular clusters.

To estimate the distance to 47~Tuc without extrapolation, we create a sample of sub-giant and giant stars and measure their proper motions using ultraviolet images in the core where the effects of rotation and anisotropy are minimized. Furthermore, this sample is only slightly less massive ($1.5\%$) than the stars for which the radial velocity dispersion has been measured, so the correction for the measured mass segregation is less than one percent. Our distance of $4.32\pm0.47$~kpc is larger than previous kinematic estimates \citep{2006ApJS..166..249M,2015ApJ...803...29W} because we have attempted to restrict our kinematic sample to stars very close in mass to the stars observed for the radial velocities.  We could achieve this by using photometry in the ultraviolet where even a portion of the giant branch is not saturated, so precision astrometry is possible. Because we expect the stellar velocities to decrease with increasing mass, using a more massive sample for the proper motion yields a larger distance estimate.  This reduces the tension with the distance of 4.7~kpc that we have adopted in this paper as determined from white dwarfs \citep{2012AJ....143...50W}.  In fact this new kinematic distance agrees with the white dwarf distance within the ninety-percent confidence region.  The agreement with the distance determined from main-sequence fitting \citep{2007A&A...476..243S,2009AJ....138.1455B}, the tip of the red-giant branch \citep{2008ApJ...686L..87B} and the analysis of eclipsing binaries in the cluster \citep{2007AJ....134..541K,2010AJ....139..329T}.

\begin{acknowledgements}
The research discussed is based on NASA/ESA Hubble Space Telescope observations obtained at the Space Telescope Science Institute, which is operated by the Association of Universities for Research in Astronomy Inc. under NASA contract NAS5-26555. These observations are associated with proposals GO-11677 and GO-12971 (PI: Richer). This work was supported by NASA/HST grants GO-11677 and GO-12971, the Natural Sciences and Engineering Research Council of Canada, the Canadian Foundation for Innovation, the British Columbia Knowledge Development Fund.  It has made used of the NASA ADS, arXiv.org and the Mikulski Archive for Space Telescopes (MAST).
\end{acknowledgements}

\bibliography{47tuc}

\begin{thebibliography}{64}
\expandafter\ifx\csname natexlab\endcsname\relax\def\natexlab#1{#1}\fi

\bibitem[{{Anderson} \& {King}(1996)}]{1996ASPC...92..257A}
{Anderson}, J. \& {King}, I.~R. 1996, in Astronomical Society of the Pacific
  Conference Series, Vol.~92, Formation of the Galactic Halo...Inside and Out,
  ed. H.~L. {Morrison} \& A.~{Sarajedini}, 257

\bibitem[{{Anderson} \& {King}(2003)}]{2003AJ....126..772A}
{Anderson}, J. \& {King}, I.~R. 2003, \aj, 126, 772

\bibitem[{{Anderson} {et~al.}(2008){Anderson}, {King}, {Richer}, {Fahlman},
  {Hansen}, {Hurley}, {Kalirai}, {Rich}, \& {Stetson}}]{2008AJ....135.2114A}
{Anderson}, J., {King}, I.~R., {Richer}, H.~B., {Fahlman}, G.~G., {Hansen},
  B.~M.~S., {Hurley}, J., {Kalirai}, J.~S., {Rich}, R.~M., \& {Stetson}, P.~B.
  2008, \aj, 135, 2114

\bibitem[{{Baraffe} {et~al.}(2015){Baraffe}, {Homeier}, {Allard}, \&
  {Chabrier}}]{2015A&A...577A..42B}
{Baraffe}, I., {Homeier}, D., {Allard}, F., \& {Chabrier}, G. 2015, \aap, 577,
  A42

\bibitem[{{Bellini} {et~al.}(2014){Bellini}, {Anderson}, {van der Marel},
  {Watkins}, {King}, {Bianchini}, {Chanam{\'e}}, {Chandar}, {Cool}, {Ferraro},
  {Ford}, \& {Massari}}]{2014ApJ...797..115B}
{Bellini}, A., {Anderson}, J., {van der Marel}, R.~P., {Watkins}, L.~L.,
  {King}, I.~R., {Bianchini}, P., {Chanam{\'e}}, J., {Chandar}, R., {Cool},
  A.~M., {Ferraro}, F.~R., {Ford}, H., \& {Massari}, D. 2014, \apj, 797, 115

\bibitem[{{Bellini} {et~al.}(2017){Bellini}, {Bianchini}, {Varri}, {Anderson},
  {Piotto}, {van der Marel}, {Vesperini}, \& {Watkins}}]{2017ApJ...844..167B}
{Bellini}, A., {Bianchini}, P., {Varri}, A.~L., {Anderson}, J., {Piotto}, G.,
  {van der Marel}, R.~P., {Vesperini}, E., \& {Watkins}, L.~L. 2017, \apj, 844,
  167

\bibitem[{{Bergbusch} \& {Stetson}(2009)}]{2009AJ....138.1455B}
{Bergbusch}, P.~A. \& {Stetson}, P.~B. 2009, \aj, 138, 1455

\bibitem[{{Bianchini} {et~al.}(2016){Bianchini}, {van de Ven}, {Norris},
  {Schinnerer}, \& {Varri}}]{2016MNRAS.458.3644B}
{Bianchini}, P., {van de Ven}, G., {Norris}, M.~A., {Schinnerer}, E., \&
  {Varri}, A.~L. 2016, \mnras, 458, 3644

\bibitem[{{Bianchini} {et~al.}(2013){Bianchini}, {Varri}, {Bertin}, \&
  {Zocchi}}]{2013ApJ...772...67B}
{Bianchini}, P., {Varri}, A.~L., {Bertin}, G., \& {Zocchi}, A. 2013, \apj, 772,
  67

\bibitem[{{Bono} {et~al.}(2008){Bono}, {Stetson}, {Sanna}, {Piersimoni},
  {Freyhammer}, {Bouzid}, {Buonanno}, {Calamida}, {Caputo}, {Corsi}, {Di
  Cecco}, {Dall'Ora}, {Ferraro}, {Iannicola}, {Monelli}, {Nonino}, {Pulone},
  {Sterken}, {Storm}, {Tuvikene}, \& {Walker}}]{2008ApJ...686L..87B}
{Bono}, G., {Stetson}, P.~B., {Sanna}, N., {Piersimoni}, A., {Freyhammer},
  L.~M., {Bouzid}, Y., {Buonanno}, R., {Calamida}, A., {Caputo}, F., {Corsi},
  C.~E., {Di Cecco}, A., {Dall'Ora}, M., {Ferraro}, I., {Iannicola}, G.,
  {Monelli}, M., {Nonino}, M., {Pulone}, L., {Sterken}, C., {Storm}, J.,
  {Tuvikene}, T., \& {Walker}, A.~R. 2008, \apjl, 686, L87

\bibitem[{Brys {et~al.}(2004)Brys, Hubert, \&
  Struyf}]{doi:10.1198/106186004X12632}
Brys, G., Hubert, M., \& Struyf, A. 2004, Journal of Computational and
  Graphical Statistics, 13, 996

\bibitem[{{Cardelli} {et~al.}(1989){Cardelli}, {Clayton}, \&
  {Mathis}}]{1989ApJ...345..245C}
{Cardelli}, J.~A., {Clayton}, G.~C., \& {Mathis}, J.~S. 1989, \apj, 345, 245

\bibitem[{{Cioni} {et~al.}(2016){Cioni}, {Bekki}, {Girardi}, {de Grijs},
  {Irwin}, {Ivanov}, {Marconi}, {Oliveira}, {Piatti}, {Ripepi}, \& {van
  Loon}}]{2016A&A...586A..77C}
{Cioni}, M.-R.~L., {Bekki}, K., {Girardi}, L., {de Grijs}, R., {Irwin}, M.~J.,
  {Ivanov}, V.~D., {Marconi}, M., {Oliveira}, J.~M., {Piatti}, A.~E., {Ripepi},
  V., \& {van Loon}, J.~T. 2016, \aap, 586, A77

\bibitem[{{Da Costa}(1982)}]{1982AJ.....87..990D}
{Da Costa}, G.~S. 1982, \aj, 87, 990

\bibitem[{{Davoust}(1986)}]{1986A&A...166..177D}
{Davoust}, E. 1986, \aap, 166, 177

\bibitem[{{Feast} \& {Thackeray}(1960)}]{1960MNRAS.120..463F}
{Feast}, M.~W. \& {Thackeray}, A.~D. 1960, \mnras, 120, 463

\bibitem[{{Gebhardt} {et~al.}(1995){Gebhardt}, {Pryor}, {Williams}, \&
  {Hesser}}]{1995AJ....110.1699G}
{Gebhardt}, K., {Pryor}, C., {Williams}, T.~B., \& {Hesser}, J.~E. 1995, \aj,
  110, 1699

\bibitem[{Goldsbury {et~al.}(2013)Goldsbury, Heyl, \&
  Richer}]{2013arXiv1308.3706G}
Goldsbury, R., Heyl, J., \& Richer, H. 2013, \apj, 778, 57

\bibitem[{{Goldsbury} {et~al.}(2010){Goldsbury}, {Richer}, {Anderson},
  {Dotter}, {Sarajedini}, \& {Woodley}}]{2010AJ....140.1830G}
{Goldsbury}, R., {Richer}, H.~B., {Anderson}, J., {Dotter}, A., {Sarajedini},
  A., \& {Woodley}, K. 2010, \aj, 140, 1830

\bibitem[{{Goldsbury} {et~al.}(2011){Goldsbury}, {Richer}, {Anderson},
  {Dotter}, {Sarajedini}, \& {Woodley}}]{2011AJ....142...66G}
---. 2011, \aj, 142, 66

\bibitem[{{Harris}(1996)}]{1996AJ....112.1487H}
{Harris}, W.~E. 1996, \aj, 112, 1487, (2010 edition)

\bibitem[{{Heinke} {et~al.}(2005){Heinke}, {Grindlay}, {Edmonds}, {Cohn},
  {Lugger}, {Camilo}, {Bogdanov}, \& {Freire}}]{2005ApJ...625..796H}
{Heinke}, C.~O., {Grindlay}, J.~E., {Edmonds}, P.~D., {Cohn}, H.~N., {Lugger},
  P.~M., {Camilo}, F., {Bogdanov}, S., \& {Freire}, P.~C. 2005, \apj, 625, 796

\bibitem[{Heyl {et~al.}(2015{\natexlab{a}})Heyl, Kalirai, Richer, Marigo,
  Antolini, Goldsbury, \& Parada}]{Heyl14massloss}
Heyl, J., Kalirai, J., Richer, H.~B., Marigo, P., Antolini, E., Goldsbury, R.,
  \& Parada, J. 2015{\natexlab{a}}, \apj, 810, 127 (8 pages)

\bibitem[{Heyl {et~al.}(2015{\natexlab{b}})Heyl, Richer, Antolini, Goldsbury,
  Kalirai, Parada, \& Tremblay}]{Heyl14diff}
Heyl, J., Richer, H.~B., Antolini, E., Goldsbury, R., Kalirai, J., Parada, J.,
  \& Tremblay, P.-E. 2015{\natexlab{b}}, \apj, 804, 53, (12 pages)

\bibitem[{Heyl {et~al.}(2012)Heyl, Richer, Anderson, Fahlman, Dotter, Hurley,
  Kalirai, Rich, Shara, Stetson, Woodley, \& Zurek}]{Heyl116397dyn}
Heyl, J.~S., Richer, H., Anderson, J., Fahlman, G., Dotter, A., Hurley, J.,
  Kalirai, J., Rich, R.~M., Shara, M., Stetson, P., Woodley, K.~H., \& Zurek,
  D. 2012, \apj, 761, 51 (25 pages)

\bibitem[{{Jedrzejewski}(1987)}]{1987MNRAS.226..747J}
{Jedrzejewski}, R.~I. 1987, \mnras, 226, 747

\bibitem[{{Kalirai} {et~al.}(2008){Kalirai}, {Hansen}, {Kelson}, {Reitzel},
  {Rich}, \& {Richer}}]{2008ApJ...676..594K}
{Kalirai}, J.~S., {Hansen}, B.~M.~S., {Kelson}, D.~D., {Reitzel}, D.~B.,
  {Rich}, R.~M., \& {Richer}, H.~B. 2008, \apj, 676, 594

\bibitem[{{Kalirai} {et~al.}(2012){Kalirai}, {Richer}, {Anderson}, {Dotter},
  {Fahlman}, {Hansen}, {Hurley}, {King}, {Reitzel}, {Rich}, {Shara}, {Stetson},
  \& {Woodley}}]{2012AJ....143...11K}
{Kalirai}, J.~S., {Richer}, H.~B., {Anderson}, J., {Dotter}, A., {Fahlman},
  G.~G., {Hansen}, B.~M.~S., {Hurley}, J., {King}, I.~R., {Reitzel}, D.,
  {Rich}, R.~M., {Shara}, M.~M., {Stetson}, P.~B., \& {Woodley}, K.~A. 2012,
  \aj, 143, 11

\bibitem[{{Kaluzny} {et~al.}(2007){Kaluzny}, {Thompson}, {Rucinski}, {Pych},
  {Stachowski}, {Krzeminski}, \& {Burley}}]{2007AJ....134..541K}
{Kaluzny}, J., {Thompson}, I.~B., {Rucinski}, S.~M., {Pych}, W., {Stachowski},
  G., {Krzeminski}, W., \& {Burley}, G.~S. 2007, \aj, 134, 541

\bibitem[{{King}(1966)}]{1966AJ.....71...64K}
{King}, I.~R. 1966, \aj, 71, 64

\bibitem[{{Ku{\v c}inskas} {et~al.}(2014){Ku{\v c}inskas}, {Dobrovolskas}, \&
  {Bonifacio}}]{2014A&A...568L...4K}
{Ku{\v c}inskas}, A., {Dobrovolskas}, V., \& {Bonifacio}, P. 2014, \aap, 568,
  L4

\bibitem[{{Lane} {et~al.}(2010){Lane}, {Brewer}, {Kiss}, {Lewis}, {Ibata},
  {Siebert}, {Bedding}, {Sz{\'e}kely}, \& {Szab{\'o}}}]{2010ApJ...711L.122L}
{Lane}, R.~R., {Brewer}, B.~J., {Kiss}, L.~L., {Lewis}, G.~F., {Ibata}, R.~A.,
  {Siebert}, A., {Bedding}, T.~R., {Sz{\'e}kely}, P., \& {Szab{\'o}}, G.~M.
  2010, \apjl, 711, L122

\bibitem[{{Leonard} \& {Merritt}(1989)}]{1989ApJ...3339..195L}
{Leonard}, P.~J.~T. \& {Merritt}, D. 1989, \apj, 339, 195

\bibitem[{{Lupton}(1993)}]{1993stp..book.....L}
{Lupton}, R. 1993, {Statistics in theory and practice} (Princeton, N.J.:
  Princeton University Press)

\bibitem[{{Lupton} \& {Gunn}(1987)}]{1987AJ.....93.1106L}
{Lupton}, R.~H. \& {Gunn}, J.~E. 1987, \aj, 93, 1106

\bibitem[{{Mayor} {et~al.}(1984){Mayor}, {Benz}, {Imbert}, {Martin}, {Prevot},
  {Andersen}, {Nordstrom}, {Ardeberg}, {Lindgren}, \&
  {Maurice}}]{1984A&A...134..118M}
{Mayor}, M., {Benz}, W., {Imbert}, M., {Martin}, N., {Prevot}, L., {Andersen},
  J., {Nordstrom}, B., {Ardeberg}, A., {Lindgren}, H., \& {Maurice}, E. 1984,
  \aap, 134, 118

\bibitem[{{Mayor} {et~al.}(1983){Mayor}, {Imbert}, {Andersen}, {Ardeberg},
  {Baranne}, {Benz}, {Ischi}, {Lindgren}, {Martin}, {Maurice}, {Nordstrom}, \&
  {Prevot}}]{1983A&AS...54..495M}
{Mayor}, M., {Imbert}, M., {Andersen}, J., {Ardeberg}, A., {Baranne}, A.,
  {Benz}, W., {Ischi}, E., {Lindgren}, H., {Martin}, N., {Maurice}, E.,
  {Nordstrom}, B., \& {Prevot}, L. 1983, \aaps, 54, 495

\bibitem[{{McLaughlin} {et~al.}(2006){McLaughlin}, {Anderson}, {Meylan},
  {Gebhardt}, {Pryor}, {Minniti}, \& {Phinney}}]{2006ApJS..166..249M}
{McLaughlin}, D.~E., {Anderson}, J., {Meylan}, G., {Gebhardt}, K., {Pryor}, C.,
  {Minniti}, D., \& {Phinney}, S. 2006, \apjs, 166, 249, (MAM06)

\bibitem[{{McLaughlin} \& {van der Marel}(2005)}]{2005ApJS..161..304M}
{McLaughlin}, D.~E. \& {van der Marel}, R.~P. 2005, \apjs, 161, 304

\bibitem[{{Meylan}(1988)}]{1988A&A...191..215M}
{Meylan}, G. 1988, \aap, 191, 215

\bibitem[{{Meylan}(1989)}]{1989A&A...214..106M}
---. 1989, \aap, 214, 106

\bibitem[{{Meylan} \& {Mayor}(1986)}]{1986A&A...166..122M}
{Meylan}, G. \& {Mayor}, M. 1986, \aap, 166, 122

\bibitem[{{Michie}(1963{\natexlab{a}})}]{1963MNRAS.125..127M}
{Michie}, R.~W. 1963{\natexlab{a}}, \mnras, 125, 127

\bibitem[{{Michie}(1963{\natexlab{b}})}]{1963MNRAS.126..499M}
---. 1963{\natexlab{b}}, \mnras, 126, 499

\bibitem[{Parada {et~al.}(2016)Parada, Richer, Heyl, Kalirai, \&
  Goldsbury}]{2016arXiv160505740P}
Parada, J., Richer, H., Heyl, J., Kalirai, J., \& Goldsbury, R. 2016, \apj,
  826, 88

\bibitem[{{Parada} {et~al.}(2016){Parada}, {Richer}, {Heyl}, {Kalirai}, \&
  {Goldsbury}}]{2016arXiv160902115P}
{Parada}, J., {Richer}, H., {Heyl}, J., {Kalirai}, J., \& {Goldsbury}, R. 2016,
  \apj, accepted

\bibitem[{{Paxton} {et~al.}(2011){Paxton}, {Bildsten}, {Dotter}, {Herwig},
  {Lesaffre}, \& {Timmes}}]{2011ApJS..192....3P}
{Paxton}, B., {Bildsten}, L., {Dotter}, A., {Herwig}, F., {Lesaffre}, P., \&
  {Timmes}, F. 2011, \apjs, 192, 3

\bibitem[{{Ratnatunga} \& {Bahcall}(1985)}]{1985ApJS...59...63R}
{Ratnatunga}, K.~U. \& {Bahcall}, J.~N. 1985, \apjs, 59, 63

\bibitem[{Richer {et~al.}(2013{\natexlab{a}})Richer, Heyl, Anderson, Kalirai,
  Shara, Fahlman, \& Rich}]{Rich1347tuc}
Richer, H., Heyl, J., Anderson, J., Kalirai, J.~S., Shara, M., Fahlman, G., \&
  Rich, R.~M. 2013{\natexlab{a}}, \apjl, 771, L15

\bibitem[{Richer {et~al.}(2013{\natexlab{b}})Richer, Goldsbury, Heyl, Hurley,
  Dotter, Kalirai, Woodley, Fahlman, Rich, \& Shara}]{2013ApJ...778..104R}
Richer, H.~B., Goldsbury, R., Heyl, J., Hurley, J., Dotter, A., Kalirai, J.,
  Woodley, K., Fahlman, G., Rich, R., \& Shara, M. 2013{\natexlab{b}}, \apj,
  778, 104

\bibitem[{Rousseeuw \& Croux(1993)}]{Rous93}
Rousseeuw, P.~J. \& Croux, C. 1993, Journal of the American Statistical
  Association, 88, 1273

\bibitem[{{Salaris} {et~al.}(2016){Salaris}, {Cassisi}, \&
  {Pietrinferni}}]{2016A&A...590A..64S}
{Salaris}, M., {Cassisi}, S., \& {Pietrinferni}, A. 2016, \aap, 590, A64

\bibitem[{{Salaris} {et~al.}(2007){Salaris}, {Held}, {Ortolani}, {Gullieuszik},
  \& {Momany}}]{2007A&A...476..243S}
{Salaris}, M., {Held}, E.~V., {Ortolani}, S., {Gullieuszik}, M., \& {Momany},
  Y. 2007, \aap, 476, 243

\bibitem[{{Sirianni} {et~al.}(2005)}]{2005PASP..117.1049S}
{Sirianni}, M. {et~al.} 2005, \pasp, 117, 1049

\bibitem[{{Spitzer} \& {Hart}(1971)}]{1971ApJ...164..399S}
{Spitzer}, Jr., L. \& {Hart}, M.~H. 1971, \apj, 164, 399

\bibitem[{{Thompson} {et~al.}(2010){Thompson}, {Kaluzny}, {Rucinski},
  {Krzeminski}, {Pych}, {Dotter}, \& {Burley}}]{2010AJ....139..329T}
{Thompson}, I.~B., {Kaluzny}, J., {Rucinski}, S.~M., {Krzeminski}, W., {Pych},
  W., {Dotter}, A., \& {Burley}, G.~S. 2010, \aj, 139, 329

\bibitem[{{Trager} {et~al.}(1995){Trager}, {King}, \&
  {Djorgovski}}]{1995AJ....109..218T}
{Trager}, S.~C., {King}, I.~R., \& {Djorgovski}, S. 1995, \aj, 109, 218

\bibitem[{{Varri} \& {Bertin}(2012)}]{2012A&A...540A..94V}
{Varri}, A.~L. \& {Bertin}, G. 2012, \aap, 540, A94

\bibitem[{{Ventura} {et~al.}(2014){Ventura}, {Criscienzo}, {D'Antona},
  {Vesperini}, {Tailo}, {Dell'Agli}, \& {D'Ercole}}]{2014MNRAS.437.3274V}
{Ventura}, P., {Criscienzo}, M.~D., {D'Antona}, F., {Vesperini}, E., {Tailo},
  M., {Dell'Agli}, F., \& {D'Ercole}, A. 2014, \mnras, 437, 3274

\bibitem[{{Walker} {et~al.}(2006){Walker}, {Mateo}, {Olszewski}, {Bernstein},
  {Wang}, \& {Woodroofe}}]{2006AJ....131.2114W}
{Walker}, M.~G., {Mateo}, M., {Olszewski}, E.~W., {Bernstein}, R., {Wang}, X.,
  \& {Woodroofe}, M. 2006, \aj, 131, 2114

\bibitem[{{Watkins} {et~al.}(2015{\natexlab{a}}){Watkins}, {van der Marel},
  {Bellini}, \& {Anderson}}]{2015ApJ...803...29W}
{Watkins}, L.~L., {van der Marel}, R.~P., {Bellini}, A., \& {Anderson}, J.
  2015{\natexlab{a}}, \apj, 803, 29

\bibitem[{{Watkins} {et~al.}(2015{\natexlab{b}}){Watkins}, {van der Marel},
  {Bellini}, \& {Anderson}}]{2015ApJ...812..149W}
---. 2015{\natexlab{b}}, \apj, 812, 149

\bibitem[{Woodley {et~al.}(2012)Woodley, Goldsbury, Kalirai, Richer, Tremblay,
  Anderson, Bergeron, Dotter, Esteves, Fahlman, Hansen, Heyl, Hurley, Rich,
  Shara, \& Stetson}]{2012AJ....143...50W}
Woodley, K.~A., Goldsbury, R., Kalirai, J.~S., Richer, H.~B., Tremblay, P.-E.,
  Anderson, J., Bergeron, P., Dotter, A., Esteves, L., Fahlman, G.~G., Hansen,
  B. M.~S., Heyl, J., Hurley, J., Rich, R.~M., Shara, M.~M., \& Stetson, P.~B.
  2012, \aj, 143, 50

\bibitem[{{Zhang} {et~al.}(2015){Zhang}, {Li}, {de Grijs}, {Bekki}, {Deng},
  {Zaggia}, {Rubele}, {Piatti}, {Cioni}, {Emerson}, {For}, {Ripepi}, {Marconi},
  {Ivanov}, \& {Chen}}]{2015ApJ...815...95Z}
{Zhang}, C., {Li}, C., {de Grijs}, R., {Bekki}, K., {Deng}, L., {Zaggia}, S.,
  {Rubele}, S., {Piatti}, A.~E., {Cioni}, M.-R.~L., {Emerson}, J., {For},
  B.-Q., {Ripepi}, V., {Marconi}, M., {Ivanov}, V.~D., \& {Chen}, L. 2015,
  \apj, 815, 95

\end{thebibliography}
\bibliographystyle{apj}

\appendix
\paragraph{Analytic results for anisotropic phase space distributions}

We can make substantial progress finding closed-form
expressions using the Michie-King model or anisotropic
lowered isothermal profile
\citep{1963MNRAS.125..127M,1966AJ.....71...64K}
\begin{equation}
f = 
\rho_1(2\pi\sigma^2)^{-3/2} e^{-J^2/(2r_a^2\sigma^2)} \left ( e^{\epsilon/\sigma^2}- 1 \right )
\label{eq:20}
\end{equation}
if $\epsilon = \Psi - \frac{1}{2} v^2>0$ and zero otherwise; $\Psi$ is
the gravitational potential, $\sigma$ is a characteristic velocity
dispersion and $\rho_1$ is a characteristic density.  

Let us define slightly more general distribution function as follows
\begin{equation}
f' = \frac{\d N}{\d^3 x \d^3 u} = 
\rho_1 (\pi)^{-3/2} e^{-\alpha u_t^2 - \gamma u_r^2} \left ( e^{u_0^2 -
    u_r^2 - u_t^2}- 1 \right )
\label{eq:21}
\end{equation}
where we have made the following changes of variables,
$u_0^2=\Psi/\sigma^2$, $u_r^2 + u_t^2 = v^2/(2\sigma^2)$ and
$\alpha=r^2/r_a^2$.

Integrating this
distribution function over velocity yields the density distribution
\begin{eqnarray}
\rho &=& \frac{\rho_1}{(\alpha+1)(\alpha-\gamma)} \Biggr [ \mathrm{erf}\left(u_0
    \sqrt{\gamma}\right) \frac{(\alpha+1) \gamma -\alpha\sqrt{\gamma} +
    \alpha^2}{\alpha\sqrt{\gamma}}+ 
%\nonumber \\
%& & 
%~~~ 
\mathrm{erf} \left ( u_0 \sqrt{\gamma+1} \right ) e^{\alpha u_0^2}
\frac{\alpha-\gamma}{\sqrt{\gamma+1}  } + \label{eq:22} 
\\
& & 
~~~ 
 2 F \left ( u_0 \sqrt{\alpha-\gamma} \right ) e^{-\gamma u_0^2}
\frac{\sqrt{\alpha-\gamma}}{\alpha\sqrt{\pi}} 
\Biggr ] \nonumber
\end{eqnarray}
where $F(z)$  is Dawson's integral,
\begin{equation}
F(z) = 	e^{-z^2} \int_0^ze^{y^2} dy.
\label{eq:23}
\end{equation}

Although this distribution function is more general than the original
distribution function (Eq.~\ref{eq:20}), the resulting density has the
following very useful properties,
\begin{equation}~
\langle v_t^2 \rangle = -2\sigma^2 \left ( \frac{d \ln \rho}{d \alpha}
\right ) \textrm{and}~
\langle v_r^2 \rangle = -2\sigma^2 \left ( \frac{d \ln \rho}{d \gamma}
\right ).
\label{eq:24}\end{equation}
Furthermore, all of the even moments can be expressed in terms of
derivatives of the density; the odd moments vanish by symmetry.

For example if we calculate these values for $\alpha=\gamma=0$, we
obtain the classic results
\begin{equation}
\frac{\d \rho}{d \alpha} = 2 \frac{d \rho}{d \gamma} =
\frac{2 u_0}{\sqrt{\pi}} \left ( 1 + \frac{2}{3} u_0^2 + \frac{4}{15}
  u_0^4 \right ) - e^{u_0^2} \mathrm{erf} (u_0).
\label{eq:25}
\end{equation}
The one-dimensional radial velocity dispersion is one-half of the
two-dimensional tangential velocity dispersion, so the velocity
distribution is isotropic.  We recover the results for
Eq.~(\ref{eq:20}) by substituting for $u$ and $\alpha=r^2/r_a^2$,
\begin{eqnarray}
\rho &=& \frac{\rho_1 r_a^2}{r^2+r_a^2} 
\Biggr \{ \mathrm{erf} \left (\frac{\sqrt{\Psi}}{\sigma}\right )
  e^{\Psi/\sigma^2} +
%\nonumber \\* 
%& & 
 \frac{2}{\sqrt{\pi}} \left [ \frac{r_a^3}{r^3} F\left (\frac{r}{r_a}
  \frac{\sqrt{\Psi}}{\sigma}\right ) 
- \frac{\sqrt{\Psi}}{\sigma} \left ( 1 + \frac{r_a^2}{r^2} \right ) \right ]
\Biggr  \}. 
\label{eq:26}
\end{eqnarray}
where $F(z)$ is Dawson's integral.

In the limit of $r_a\rightarrow\infty$ taking the first two terms in
the Taylor series for Dawson's integral recovers the standard King
model
\begin{equation}
 F(z) =
 z - \frac{2}{3} z^3 + \frac{4}{15} z^5 + {\cal O} \left ( z^7
  \right ) .
\label{eq:27}
\end{equation} 

Furthermore, we can express the velocity moments in terms of Dawson's
integral as well
\begin{eqnarray}
\rho \langle v_r^2 \rangle &=& \frac{\rho_1 \sigma^2 r_a^2}{r^2+r_a^2} 
\Biggr \{  \mathrm{erf} \left (\frac{\sqrt{\Psi}}{\sigma}\right )
  e^{\Psi/\sigma^2} - 2 \frac{r_a^5}{r^5} \frac{1}{\sqrt{\pi}}  F\left (\frac{r}{r_a}
  \frac{\sqrt{\Psi}}{\sigma}\right )
%\nonumber \\* 
%& & 
 + 2 \frac{\sqrt{\Psi}}{\sigma}
  \frac{1}{\sqrt{\pi}} \left [ \frac{r_a^4 - r^4}{r^4}
-\frac{2}{3} \frac{\Psi}{\sigma^2} \frac{r^2+r_a^2}{r^2} \right ] 
\Biggr  \}. 
\label{eq:28}
\end{eqnarray}
and
\begin{eqnarray}
\rho \langle v_t^2 \rangle &=& \frac{\rho_1 \sigma^2 r_a^4}{\left(
    r^2+r_a^2 \right)^2} 
\Biggr \{ 2 \mathrm{erf} \left (\frac{\sqrt{\Psi}}{\sigma}\right )
  e^{\Psi/\sigma^2} +
%\nonumber \\* 
%& & 
 \frac{1}{\sqrt{\pi}}  F\left (\frac{r}{r_a} 
  \frac{\sqrt{\Psi}}{\sigma}\right ) \left [ 2 \frac{r_a}{r} \frac{\Psi}{\sigma^2} \frac{r_a^2+r^2}{r^2} +
  \frac{r_a^3}{r^3} \frac{5r^2 + 3 r_a^2}{r^2} \right ] 
\nonumber \\* 
& & 
- \frac{1}{\sqrt{\pi}} \frac{\sqrt{\Psi}}{\sigma} \left (\frac{3
    r_a^4}{r^4} + \frac{5 r_a^2}{r^2} + 2 \right )
\Biggr  \}. 
\label{eq:29}
\end{eqnarray}

\paragraph{Analytic results for rotating, anisotropic phase space distributions}
\label{sec:rotating-models}
\paragraph{Isopleths from catalogues}
\label{sec:isopl-from-catal}

An isopleth is a contour of constant density.  Rather than create
images from the star catalogues, we use the catalogues themselves to
generate isopleths.  If we limit ourselves to elliptical isopleths we
can follow the following iterative procedure, 
\begin{eqnarray*}
x_0 &\rightarrow& x_0 + 0.2 \frac{1}{N} \sum (x_i - x_0),~ 
y_0 \rightarrow y_0 + 0.2 \frac{1}{N} \sum (y_i - y_0),~  
%\\
a \rightarrow F a,~ 
b \rightarrow F^{-1} b \textrm{~and~}
% \\
\theta \rightarrow  \theta + \Delta \theta
\end{eqnarray*}
% \begin{eqnarray*}
% x_0 &\rightarrow& x_0 + 0.2 \frac{1}{N} \sum (x_i - x_0)
% %\\
% y_0 &\rightarrow& y_0 + 0.2 \frac{1}{N} \sum (y_i - y_0) 
% %\\
% a &\rightarrow& F a 
% \\
% b &\rightarrow& F^{-1} b \\
% \theta &\rightarrow & \theta + \Delta \theta
% \end{eqnarray*}
where
\begin{equation}
F = \exp \left \{ 0.5 \sum_i \frac{1}{r_i^2} \left [ 2 \frac{\delta x_R^2}{r_i^2} - 1 \right ] \left [ \sum_i
  \frac{1}{r_i^2} \right ]^{-1} \right \},~ 
%\end{equation}
%\begin{equation}
\Delta \theta = 0.05 \sum_i \frac{1}{r_i^2} \left [ 2 \frac{\delta x_R
    \delta y_R}{r_i^2} \right ] \left [ \sum_i
  \frac{1}{r_i^2} \right ]^{-1} 
\label{eq:30}\end{equation}
and 
\begin{eqnarray}
\delta x_R &=& (x_i-x_0) \cos \theta - (y_i-y_0) \sin\theta,~
% \\
\delta y_R = (x_i-x_0) \sin \theta + (y_i-y_0) \cos\theta, \\
r_i^2&=&(x_i-x_0)^2+(y_i-y_0)^2.
\label{eq:31}
\end{eqnarray}
The iteration preserves the area
enclosed of the elliptical isopleth while minimizing the variation in
the surface density along the isopleth.  This is similar in
spirit to iteration defined by \citet{1987MNRAS.226..747J}.
The summations are performed
over all stars in the catalogue such that
\begin{equation}
0.81 < \left ( \frac{\delta x_R}{a} \right )^2 + \left ( \frac{\delta y_R}{b} \right )^2 < 1.21.
\label{eq:32}
\end{equation}
A second equivalent way to define the isopleth is to find the region
that contains the most stars for a given area.  Although this seems to
be based on the global properties of the stellar distribution, by
Green's theorem this procedure results in a curve on which the surface
density is constant.  Let's calculate the number of stars over the
region
\begin{equation}
\int_D \Sigma (x,y) d x dy = \int_0^1 P(x,y) {\dot x} + Q(x,y) {\dot y} dt
\label{eq:33}\end{equation}
where $x(t)$ and $y(t)$ define the boundary of the region and
\begin{equation}
\Sigma(x,y) = \frac{\partial P}{\partial y} - \frac{\partial Q}{\partial x}.
\label{eq:34}\end{equation}
The area can also be written in terms of an integral around the
boundary
\begin{equation}
A = \frac{1}{2} \int_0^1 dt \left ( x \dot y - y \dot x \right ) .
\label{eq:35}\end{equation}
Combining these results yields the following integral to extremize
\begin{equation}
L = \int_0^1 dt \left [ P(x,y) \dot x + Q(x,y) \dot y  +
\lambda  \left ( \frac{1}{2} x \dot y - \frac{1}{2} y \dot x - A \right ) \right ].
\label{eq:36}\end{equation}
where $\lambda$ is a Lagrange multiplier and $A$ is the area of the
contour that we would like to keep fixed.  Applying Lagrange's
equation for the $x$-coordinate of the boundary yields
\begin{eqnarray}
\frac{d}{dt} \frac{\partial L}{\partial \dot x}  - \frac{\partial
  L}{\partial x} &=& 0 \label{eq:37}\\
\frac{\partial P}{\partial x} \dot x + \frac{\partial P}{\partial y} \dot y - \frac{1}{2} \lambda \dot y -
\frac{\partial P}{\partial x} \dot x
- \frac{\partial Q}{\partial x} \dot y - \frac{1}{2} \lambda \dot y &=& 0
\\
\left ( \frac{\partial P}{\partial y} - \frac{\partial Q}{\partial x}
  - \lambda \right ) \dot y &=& 0 \label{eq:38}
\end{eqnarray}
so the constant Lagrange multiplier $\lambda$ equals $\Sigma(x,y)$
along the boundary.  The boundary of a region of given area that
maximizes the number of stars within is an isopleth. Lagrange's
equations for the $y$-coordinate yield the same result.  For
elliptical isopleths the iterative procedure is typically faster and
yields smaller errors as calculated by bootstrapping the
star catalogues.  The two methods agree within their respective
confidence regions.  The second technique although more
computationally expensive is more flexible and allows general shapes
for the isopleths.

\end{document}